\def\today{\ifcase\month\or January\or
February\or March\or April\or May\or June\or
  July\or August\or September\or October\or November\or December\fi
  \space\number\day, \number\year}
\documentclass{emulateapj}

\newcommand{\TransAntiBelieve}{0.735}

\newcommand{\PlanetPeriodLong}{0.7517}
\newcommand{\PlanetPeriod}{0.752}
\newcommand{\PlanetRadius}{0.134} % //0.133904

\newcommand{\PlanetRadiusEarth}{1.50} % //1.62255
\newcommand{\PlanetInc}{86.0}
\newcommand{\PlanetChiPerc}{0.438}
\newcommand{\PlanetEphemeris}{2403.711} % 2403.711124
\newcommand{\PlanetMassEarth}{1.8} 
\newcommand{\PlanetMassJupiter}{0.006} 

\newcommand{\limbone}{0.82523168}
\newcommand{\limbtwo}{-0.97755590}
\newcommand{\limbthree}{1.6867374}
\newcommand{\limbfour}{-0.65702013}

\newcommand{\Pmin}{0.4}
\newcommand{\Pmax}{6.9}
\newcommand{\NumPassedSelection}{3}

\newcommand{\EtaChoice}{0.000219}

\newcommand{\HowMany}{100}
\newcommand{\RTrojanJupiter}{0.24}
\newcommand{\RTrojanEarth}{2.7}
\newcommand{\MTrojanEarth}{11}

\newcommand{\Rjupsmall}{0.15} % HD 189733
\newcommand{\Rjuplarge}{0.31} % HD 189733
\newcommand{\Rearthsmall}{1.7} % HD 189733
\newcommand{\Rearthlarge}{3.5} % HD 189733
\newcommand{\Rratiosmall}{0.020} % HD 189733
\newcommand{\Rratiolarge}{0.042} % HD 189733
\newcommand{\MassNeptune}{12.2} % just times massearth large by 16/30

\newcommand{\MearthsmallIce}{1.3} 
\newcommand{\MearthsmallRock}{4.8} 
\newcommand{\MearthsmallIron}{27} 
\newcommand{\MearthlargeIce}{23}

\begin{document}

\slugcomment{Draft \today}

\shorttitle{Looking for Super-Earths in the HD 189733 system}

\shortauthors{Croll et~al.}

\title{LOOKING FOR SUPER-EARTHS IN THE HD 189733 SYSTEM:
A SEARCH FOR TRANSITS IN {\it MOST}\altaffilmark{1} SPACE-BASED PHOTOMETRY}

\author{Bryce Croll\altaffilmark{2}, Jaymie M. Matthews\altaffilmark{3},
Jason F. Rowe\altaffilmark{3}, Brett Gladman \altaffilmark{3}, 
Eliza Miller-Ricci\altaffilmark{4},
Dimitar Sasselov\altaffilmark{4},
Gordon A.H. Walker\altaffilmark{5}, Rainer Kuschnig\altaffilmark{3}, Douglas N.C. Lin \altaffilmark{6}, 
David B. Guenther\altaffilmark{7}, Anthony F.J. Moffat\altaffilmark{8}, Slavek M.
Rucinski\altaffilmark{9}, Werner W. Weiss\altaffilmark{10}}

\altaffiltext{1}{Based on data from the MOST satellite, a Canadian Space
Agency mission, jointly operated by Dynacon Inc., the University of Toronto
Institute of Aerospace Studies and the University of British Columbia, with
the assistance of the University of Vienna.}

\altaffiltext{2}{Deptartment of Astronomy and Astrophysics, University of Toronto, 50 St. George Street, Toronto, ON 
M5S 3H4, Canada;
croll@astro.utoronto.ca}

\altaffiltext{3}{Deptartment of Physics \& Astronomy, University of British
Columbia, 6224 Agricultural Road, Vancouver, BC V6T 1Z1, Canada;
matthews@phas.ubc.ca, rowe@phas.ubc.ca, gladman@phas.ubc.ca,
kuschnig@phas.ubc.ca}

\altaffiltext{4}{Harvard-Smithsonian Center for Astrophysics,
60 Garden
Street, Cambridge, MA 02138, USA; emillerr@cfa.harvard.edu, sasselov@cfa.harvard.edu}

\altaffiltext{5}{1234 Hewlett Place, Victoria, BC V8S 4P7, Canada;
gordonwa@uvic.ca}

\altaffiltext{6}{University of California Observatories, Lick Observatory, University of California, Santa Cruz, CA 95064, USA;
lin@ucolick.org}

\altaffiltext{7}{Department of Astronomy and Physics, St. Mary's
University, Halifax, NS B3H 3C3, Canada; guenther@ap.stmarys.ca}

\altaffiltext{8}{D\'ept de physique, Univ de
Montr\'eal, C.P.\ 6128, Succ.\ Centre-Ville, Montr\'eal, QC H3C 3J7,
Obs du mont M\'egantic, Canada; moffat@astro.umontreal.ca}

\altaffiltext{9}{Dept. Astronomy \& Astrophysics, David Dunlap Obs., Univ.
Toronto, P.O.~Box 360, Richmond Hill, ON L4C 4Y6, Canada;
rucinski@astro.utoronto.ca}

\altaffiltext{10}{Institut f\"ur Astronomie, Universit\"at Wien,
T\"urkenschanzstrasse 17, A--1180 Wien, Austria; weiss@astro.univie.ac.at}

\begin{abstract}

We have made a comprehensive transit search for exoplanets
down to
$\simeq$1.5 - 2 Earth radii in the HD 189733 system, based on 21-days of nearly
uninterrupted broadband optical photometry obtained with the {\it MOST}
({\it Microvariability \& Oscillations of STars}) satellite in 2006.
We have searched these data for
realistic limb-darkened transits from exoplanets other than the known hot Jupiter, HD 189733b,
with periods ranging from about 0.4 days to one week. Monte Carlo statistical tests of
the data with synthetic transits inserted into the data-set allow us to rule out additional
close-in exoplanets with sizes ranging from about \Rjupsmall \ - \Rjuplarge \ $R_{J}$
(Jupiter radii), or \Rearthsmall \ - \Rearthlarge \ $R_{\earth}$ (Earth radii) on orbits
whose planes are near that of HD 189733b.
These null results constrain
theories that invoke lower-mass hot Super-Earth and hot Neptune planets in orbits similar to HD 189733b
due to the inward migration of this hot Jupiter.
This work also illustrates the feasibility of 
discovering smaller transiting planets around chromospherically active stars. 
\end{abstract}

\keywords{ planetary systems -- stars: individual: HD 189733 -- methods: data analysis -- techniques: photometric}

\setcounter{footnote}{0}      % otherwise corrupted numbers

\section{INTRODUCTION}
\label{SecIntro}

With the launch of the {\it MOST} ({\it Microvariability
\& Oscillations of STars}) (\citealt{Walker};
\citealt{Matthews}) and COROT (\citealt{Baglin};
\citealt{Barge}) satellites, and the upcoming launch 
of the Kepler (\citealt{Borucki}; \citealt{Basri})
space mission, the age of transit searches with continuous space-based
photometry is now upon us.
With these new telescopes we should now be able to utilize the transit method to probe
the terrestrial and giant-terrestrial regime
of transiting exoplanets.
We can seriously anticipate the discovery of a transiting Super-Earth \citep{ValenciaA} sized
planet orbiting a Sun-like star in the near-term future.

Numerical simulations of migrating hot Jupiters have indicated that one might
expect Earth, Super-Earth and Neptune sized planets 
in similar orbits,
or in mean-motion resonances with the hot Jupiter
(\citealt{Zhou}; \citealt{Raymond}; \citealt{Mandell}; \citealt{Fogg}).
The interior regions of these hot Jupiter exoplanetary systems, 
and specifically the mean-motion resonances of these hot Jupiters, 
have been probed with the transit-timing method in a handful of transiting exoplanetary systems
to date - these are TrES-1 (K0 V) \citep{Steffen}, HD 209458 (G0 V) (\citealt{Agol}; \citealt{Miller}),
and
HD 189733 (K0 V) (\citealt{WinnPsi}; \citealt{MillerTwo}; \citealt{PontPrep}).
The transit method,
although currently less sensitive in the mean-motion resonances of the hot Jupiter, is able to 
probe a continuous range of orbital separations in the interior regions of these hot-Jupiter systems. 
The transit method has only been used to sensitively probe the interior regions of one hot Jupiter
exoplanetary system to date; \citet{Croll} reported a null-result in the HD 209458 system from the first
space-based transit-search using nearly continuous photometry returned by the
{\it MOST} satellite.
In that work, \citet{Croll} ruled out hot Super-Earth and hot Neptune-sized planets larger than
2-4 Earth radii
in the HD 209458 system with periods ranging from 0.5 days to 2 weeks.
Here we report a search using {\it MOST} photometry and similar methods
ruling out transiting exoplanets with periods from 0.4 days to one week
in the HD 189733 system with orbital inclinations similar to that of
the known hot Jupiter exoplanet, HD 189733b.

The existence of a hot Jupiter planet, HD 189733b, in the star system HD 189733 (V = 7.67)
was first reported by \citet{Bouchy}.
The planet was
detected from radial velocity measurements in the ELODIE metallicity-biased search for transiting hot Jupiters.
Photometry of HD 189733 taken by \citet{Bouchy}
indicated the planet HD 189733b transited the star along our line of sight.
Follow-up work has determined the stellar and planetary characteristics of the detected hot Jupiter
to a high
degree of accuracy (\citealt{Bakos}; \citealt{WinnLambda}; \citealt{WinnPsi}; \citealt{Baines}).
A number of researchers have determined characteristics of the planet by
measuring the drop in flux
during the secondary eclipse. These results include the thermal emission of the planet \citep{Deming},
one of the first spectra of an extrasolar planet \citep{Grillmair},
as well as the first measurement of the day-night contrast of an extrasolar planet \citep{Knutson}.
Recent evidence indicates that HD 189733 is likely a binary system with a mid-M dwarf star, HD 189733B (period $\sim$ 3200 $yr$),
as a companion to the main K0 star, HD 189733 \citep{BakosPal}. The putative transiting planets that are
searched for in our manuscript are assumed to transit the primary star, HD 189733, 
as does the known planet HD 189733b. 
Radial-velocity measurements have already ruled out other planets than the known hot Jupiter
in this system that are more massive than approximately $\sim$32 $M_{\earth}$
(\citealt{WinnLambda} as quoted in \citealt{MillerTwo}).

Recently, HD 189733 was discovered to display quasiperiodic flux variations at the $\sim$3\%
level (\citealt{WinnPsi}; 
Matthews et al. in preparation).
% \citealt{MatthewsHD}). % PREPCHANGE
This result is in concordance with the results
of \citet{Wright}, which indicated that HD 189733 was a chromospherically active star.
Detecting planets via the transit method in chromospherically active stars is a relatively new
topic.
In addition to the original discovery of HD 189733b
around this active star, and the pioneering work of the TEP network (\citealt{Deeg}; \citealt{Doyle}),
\citet{Hebb} recently found that they could rule-out Jupiter and Neptune-sized planets
with short periods via the transit method
around the active star AU Mic.
We extend these efforts to the Neptune and Super-Earth-sized regime by investigating the continuous
and accurate photometry returned by {\it MOST} to search for
Super-Earth and Neptune sized planets with short periods around this active star. 
 
{\it MOST}'s 21-day observations of HD 189733 are presented in 
Matthews et al. (in preparation).
% \citep{MatthewsHD}. % PREPCHANGE
These observations have already been used to place limits on other hot Super-Earth and hot Neptune
exoplanets near the mean-motion resonances of the hot Jupiter
in the system through the transit-timing method \citep{MillerTwo}. We extend these efforts
here by 
ruling out hot Super-Earth and hot Neptune-sized planets using the transit method.
The rotational modulation displayed in {\it MOST}'s observations of this system has also been fit with a starspot model 
by \citet{CrollSpot}. Their model argues in favour of moderate
spin-orbit misalignment between the stellar spin axis and the orbit-normal of the hot Jupiter in this system. 

In $\S$\ref{SecMOST} the {\it MOST} photometry of HD 189733 is
described. The transit search technique is briefly summarized in
$\S$\ref{SecTrans}. The Monte Carlo statistics used to estimate the
sensitivity of the transit search are specified in $\S$\ref{SecMonte}.
The transit search routine is applied to the {\it MOST} HD 189733 data
set in $\S$\ref{SecHDTrans}, where we are able to rule-out hot Super-Earth and hot-Neptune sized planets
in this system.
Our results are summarized in 
$\S$\ref{SecDiscuss}, including a discussion of the approximate mass of exoplanets that have been ruled
out by our results for various compositions of extrasolar planets.

\section{{\it MOST} Photometry of HD 189733}
\label{SecMOST}

The {\it MOST} satellite was launched on 30 June 2003 and its initial
mission is described by \citet{Walker} and \citet{Matthews}. A
15/17.3-cm Rumak-Maksutov telescope feeds two CCDs, one originally
dedicated to tracking and the other to science, through a single,
custom, broadband filter (350 -- 700 nm). {\it MOST} was placed in
an 820-km altitude circular Sun-synchronous polar orbit with a period of
101.413 minutes. From this vantage point, {\it MOST} can monitor
stars in a continuous viewing zone (covering a declination range
$+36^{\circ} \leq \delta \leq -18^{\circ}$) within which stars can
be monitored without interruption for up to 8 weeks. Photometry of
very bright stars ($V \leq 6$) is obtained in Fabry Imaging mode, in
which a Fabry microlens projects an extended image of the telescope
pupil illuminated by the target starlight to achieve the highest
precision (see Matthews et al. 2004). Fainter stars (down to about
$V \sim 12$) can be observed in Direct Imaging mode, where 
defocused images of stars are monitored in Science CCD subrasters
(see \citealt{Rowe}).

{\it MOST} observed HD 189733 in Direct Imaging mode for 21 days during 31 July $-$ 
21 August 2006.  For the first 14 days of this run, HD 189733 was the exclusive target; 
for the final 7 days, it was shared with another Primary Science Target field during each 
101.4-min orbit of the satellite. Consequently, during the last third of the run, HD 189733 
was observed only during intervals of high stray Earthshine, resulting in somewhat 
increased photometric scatter. The duty cycles of the light curve are 94\% during the 
first 14 days and 46\% during the last 7 days.
Early in 2006 the tracking CCD system of {\it MOST} failed due to a 
particle hit. Thereafter both science and tracking were carried out with the 
Science CCD system. To avoid introducing significant tracking errors 
exposures were limited to 1.5 sec. 14 
consecutive exposures were then stacked on board to improve signal-to-noise,
and downloaded 
from the satellite every 21 sec. Approximately 112,000 of these combined observations were taken.

The photometric reduction was performed by one of us (JFR) using techniques similar 
to those described by \citet{Rowe} and
% \citet{Rowe2}. % PREPCHANGE
Rowe et al. (in preparation).
The reduction procedure of aperture 
photometry is similar to that applied to groundbased CCD photometry, but is non-differential.  
Our reduction pipeline corrects for cosmic ray hits (more frequent during satellite 
passages through the South Atlantic Anomaly [SAA]), the varying background due to 
scattered Earthshine modulated at the satellite orbital period, and flatfielding effects. 
Further details are given by 
Matthews et al. (in preparation).
% \citet{MatthewsHD}. %PREPCHANGE
It should also be noted that HD 189733, and its M-dwarf companion are
sufficiently well separated
(216 AU or 11.2'' [\citealt{BakosPal}]), that they do not contaminate the photometric signal
obtained by {\it MOST}.

\subsection{Additional selection and filtering of the data}
\label{SecMOSTReduce}

For this transit search, additional selection and filtering of the photometry was
done to optimize the data for the application of Monte Carlo statistics.
The filtering is an automated part of the transit search routine, and is thus briefly summarized
in $\S$\ref{SecTrans}. This filtering step is described in detail here.

Data obtained during passages through the SAA are conservatively
excised from the light curve, due to the increased photometric scatter
during those MOST orbital phases, without reducing seriously the phase
coverage of the exoplanetary periods searched. The transits of
the known giant planet HD 189733b are removed, at the orbital period $P$ $\sim$ 2.21857d determined by \citet{Bakos}.
The first $\sim$0.8$d$ of the data-set was excluded from the analysis due to an obvious
temperature swing-in effect common to {\it MOST}.
The obvious modulation of the light curve, due to starspot activity,
was removed with a cubic spline \citep{Press}.
The cubic spline was generated with data binned every $\sim$500 minutes;
this trend was then subtracted from the unbinned data.
This value of $\sim$500 minutes is believed to be optimum,
as it effectively balances the competing desires of having a sufficiently
short binning time to effectively remove long period modulation,
with the desire of having a long binning time to ensure that
one is able to recover 
transits with periods in the
upper range of our interesting period range ($P$ $>$ 5$d$).
Limited statistical tests have confirmed that this binning value
optimizes our sensitivity to transits of HD 189733 with periods in the upper range of our period range.
The signal to noise that an individual transit is recovered with can be highly
dependent on this binning time for transits of relatively long periods ($P$ $>$ 5$d$). 
At these longer periods the cubic-spline
can filter out much of the amplitude of the transit signal a low percentage of the time.
This effect can also be
seen with the reduced efficiency at longer periods for the ninety-degree inclination angle case of the
Monte Carlo statistics (Figure \ref{FigMonte}). 
% Because of the modulation of stray Earthshine
% with {\it MOST}'s 101.4-min orbital period, the data were phased to that
% period, and segments showing the most noticeable effects from stray
% light were also removed. The coverage in phase
% per 101.4-min {\it MOST} orbit following this cut 
% were 63\%. THIS WAS NOT DONE, WHY NOT?

{\it MOST} suffers from stray Earthshine with its 101.4-min orbital period, and this stray
light background can also be modulated at a period of 1 day and its first harmonic (due to the 
Sun-synchronous nature of the MOST satellite orbit). For this reason, 
sinusoidal fits with periods within 1\% of 1 and 0.5 d were subtracted from the data.
After these cuts, any remaining outliers
greater than $6\sigma$ were excised. 
In general, there were few points removed by this sigma-cut as few points were such extreme outliers.
It is also important to note that the magnitudes of the injected transits ($\S$\ref{SecMonte}) were always
at a level much less than this sigma-cut.
The data were also
median-subtracted, for reasons outlined in \citet{Croll}.
The automatic filtering step removed $\sim$8\% of the original data (mostly due to the
SAA and stray light corrections). 

\begin{figure}
\includegraphics[width=1.5in, height = 3.5in, angle=270]{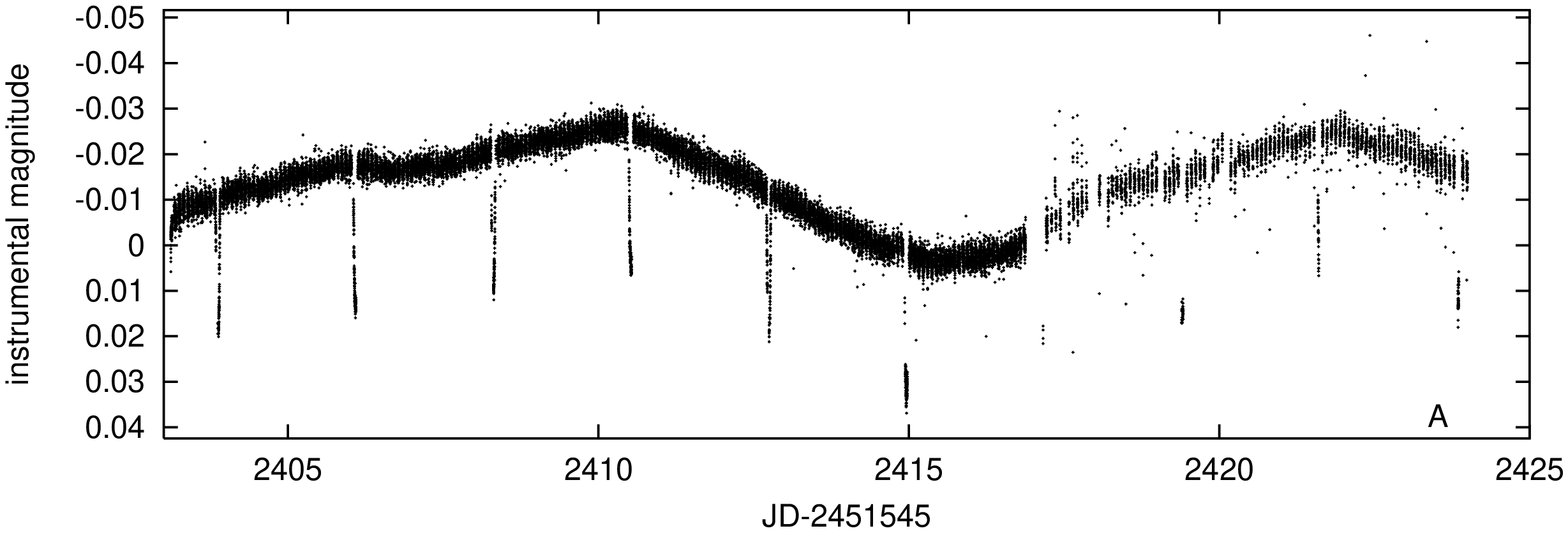}  % 2.1, 7.0
\includegraphics[width=1.5in, height = 3.5in, angle=270]{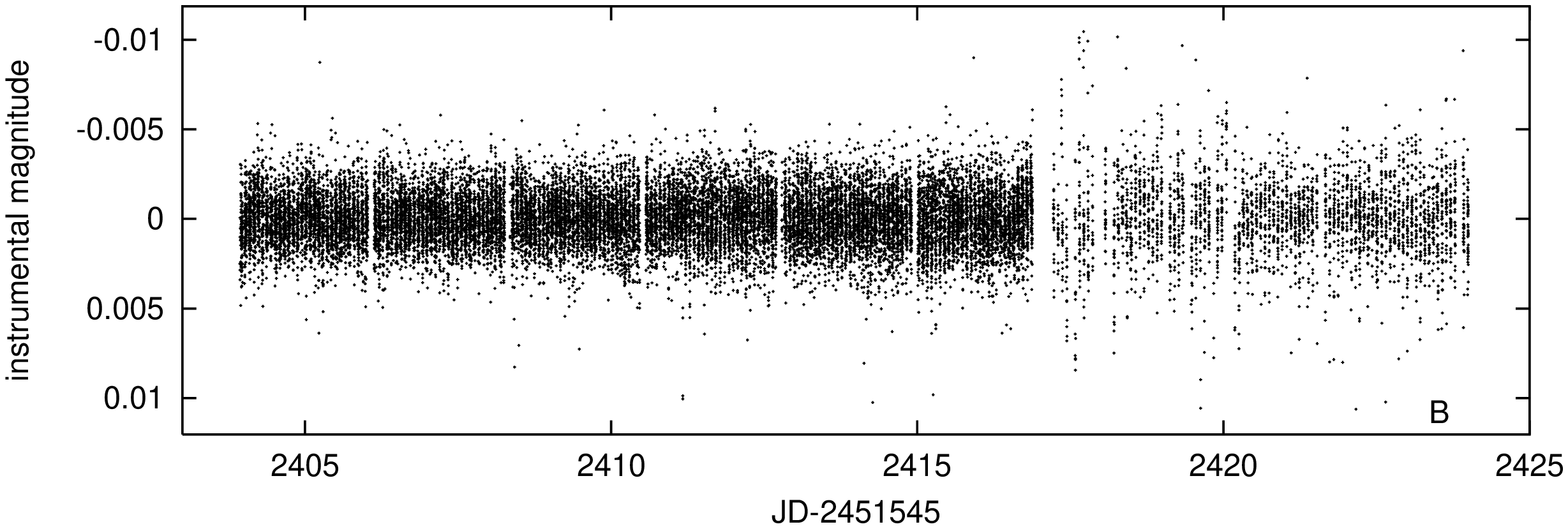}
\includegraphics[width=1.5in, height = 3.5in, angle=270]{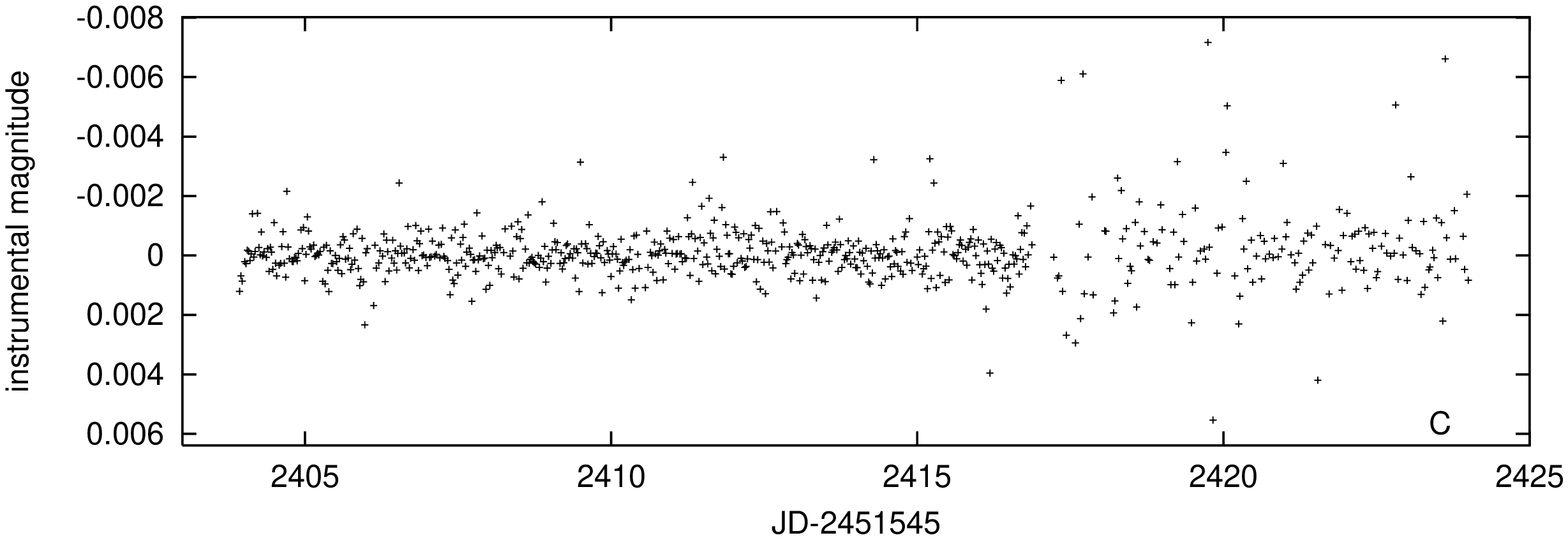}
\caption{
	{The 2006 (A) HD 189733 {\it MOST} data-set before the automated filtering step.
	The data following (B) the automatic filtering step, in which the transits of the known planet were removed, and 
	a cubic spline removed the obvious long-term variability.
	(C) The same data are shown binned in 30 minute intervals at a different vertical scale. 
	}  
	\label{FigData}
}
\end{figure}

The HD 189733 light curves before and after these reduction steps are plotted in Figure
\ref{FigData}. These filtered data (and the original reduced data) can
be downloaded from the {\it MOST} Public Data Archive at
www.astro.ubc.ca/MOST.

\section{Transit Search Algorithm}
\label{SecTrans}

The transit search routine used here is fully described in \citet{Croll}, and is briefly summarized here.
The minor modifications to the routine are highlighted below. The transit routine consists of five steps: (i) an automated filtering step (described in detail in $\S$\ref{SecMOSTReduce}),
(ii) a modified version of the EEBLS 
(Edge Effect Box-fitting Least Squares) algorithm of \citet{Kovacs} that searches for box-shaped transits, (iii) selection criteria
to select the the best candidates that can be tested for astrophysical plausibility, (iv) realistic transit-fitting
and refinement of the transit parameters, and finally (v) detection
criteria to differentiate false positives from likely transit candidates.

The detection criteria used in this study are summarized here.
We use the transit to anti-transit (the best-fit brightening event with the shape of a transit) 
ratio statistic (RS) of \citet{Burke} to quantify a believable transit.
The RS value that a transit has to exceed to be deemed a believable transit is:
$\Delta\chi^{2}\%$/$\Delta\chi^{2}_{-}\%$ $\ge$ \TransAntiBelieve, where
$\Delta\chi^{2}$ and $\Delta\chi^{2}_{-}$ refer to the normal $\chi^{2}$ (the sum of sqaured residuals)
for the best-fit transit, and anti-transit, respectively. This value
was motivated by the results of the Monte Carlo tests ($\S$\ref{SecMonte}).
The RS statistic serves to objectively quantify the significance
of the measured signal against the correlated
scatter in the data, and to thus stringently rule-out false positives.
The correlated signal to noise calculation of \citet{Pont} serves a similar purpose.
The RS will be used in this present application.  

The restriction on the quantity
f, a measure of the ratio of the strength of any one specific transit compared to the others,
as described in \citet{Croll}, and \citet{Burke}, was also relaxed such that transits were accepted
as long as f $<$ 0.85. This motivation for this less stringent criterion was based on the Monte Carlo
statistics that indicated that a more conservative 
criterion on f would rule out realistic transits, without an appreciable decrease in the number
of false positives in our data-set. 
 
The precise values used for various parameters in the EEBLS step (ii) in the transit search routine
are summarized in Table \ref{TableBls}.
We have decreased the minimum value probed by the EEBLS routine in the application of this
transit search routine to 
HD 189733, compared to its application to HD 209458 in \citet{Croll}. This is due to the conclusions
of \citet{Sahu}, who observed
a number of ultra-short period ($P$ $<$ 1.0 $d$) planet candidates in the SWEEPs survey, and noted that these planets
may preferrentially
be found around low-mass stars. HD 189733 has a stellar mass of 0.82 $M_{\odot}$ \citep{Bakos},
below the limit of 0.88 $M_{\odot}$ suggested by \citet{Sahu} to be indicative of a preference for ulta-short period planets.
The minimum period investigated by the EEBLS routine
is thus now \Pmin$d$, a value corresponding to a semi-major axis, $a_{p}$, of just under three times the stellar radius ($a_{p}$ $\approx$ 2.5$R_{*}$).
For comparison, OGLE-TR-56b has $a_{p}$$\approx$$4.4R_{*}$,
the smallest orbital radius in terms of its stellar radius of a
confirmed planet to date.
The maximum period of 6.9$d$ is chosen similar to the maximum value for HD 209458
in \citet{Croll}, as a value that would allow one to observe three distinct transits in the 21 $d$ data-set.
This period range of 0.4 to 6.9 $d$
corresponds to 0.01 to 0.07 AU.
The minimum fractional transit length is 0.75 times that of the calculated average
fractional transit length at that period - a value chosen to ensure the routine remains adequately
sensitive to transiting planets with non edge-on inclination angles. We also increase the number of bins, $Nb$,
used by the EEBLS algorithm for short periods compared to the application of \citet{Croll} to HD 209458.
The number
of bins, $Nb$, used by the EEBLS algorithm for each period, $P$, is determined by the following formula:
$[0.7\times \exp(-$P$) + 1.0]$$\times$20.0/$Qmi$, where $Qmi$ is the minimum fractional transit length to
be tested. $Qmi$, and the maximum fractional transit length, $Qma$, are determined as given in Table 
\ref{TableBls}, from our average estimate of the fractional transit length for a hypothetical planet, $Qm_{P}$,
at various periods \citep{Croll}.

In our transit selection step we now select the top three EEBLS SR signals, instead
of the top two, as well as up to two transit signals with SR signals that are
at least 1.8 times the noise floor. A full explanation of these details
is provided in \citet{Croll}.  
The stellar
characteristics used to determine the realistic limb-darkened transit model (iv) are summarized in Table \ref{TableFit}.
The precise transit parameters are refined using a Marquardt-Levenberg routine (described in detail in \citealt{Croll}).
The method used to refine our transit parameters is very similar to the Newton-Raphson method applied
by \citet{Collier} for the same purpose.

\begin{deluxetable*}{ccc}
\tabletypesize{\footnotesize}
\tablecaption{EEBLS input parameters \label{TableBls}}
\tablewidth{0pt}
\tablehead{\colhead{var}&	\colhead{definition} &	\colhead{value} \\}
\startdata
$Np$ 		& Number of Period Points searched 					& 15 000 					\\ 
$\eta$ 		& Logarithmic period step 						& \EtaChoice					\\ 
$P_{min}$	& Minimum period submitted to EEBLS algorithm				& \Pmin$d$ 					\\	
$P_{max}$	& Maximum period submitted to EEBLS algorithm				& \Pmax$d$					\\
$Qmi$		& Minimum fractional transit length to be tested 			& $0.75$$\times$$Qm_{P}$ 			\\
$Qma$ 		& Maximum fractional transit length to be tested 			& $1.1$$\times$$Qm_{P}$				\\
$Nb$ 		& Number of bins in the folded time series at each test period 		& $(0.7\times \exp(-$P$) + 1.0)$$\times$20.0/$Qmi$	\\
\enddata
\end{deluxetable*}

\begin{deluxetable}{ccc}
\tabletypesize{\footnotesize}
\tablecaption{System parameters \label{TableFit}}
\tablewidth{0pt}
\tablehead{\colhead{var}&	\colhead{definition} &	\colhead{value} \\}
\startdata
$R_{*}$ & Stellar radius & 0.755 $R_{\odot}$ \tablenotemark{a} \\
$M_{*}$ & Stellar Mass & 0.82 $M_{\odot}$ \tablenotemark{b}\\
$c_{1}$ & Non-linear limb-darkening parameter 1 &  \limbone \tablenotemark{c} \\
$c_{2}$ & Non-linear limb-darkening parameter 2 &  \limbtwo \tablenotemark{c}\\
$c_{3}$ & Non-linear limb-darkening parameter 3 &  \limbthree \tablenotemark{c}\\
$c_{4}$ & Non-linear limb-darkening parameter 4 &  \limbfour \tablenotemark{c}\\
\enddata
\tablenotetext{a}{Parameter obtained from \citet{PontPrep}}
\tablenotetext{b}{Parameter obtained from \citet{Bakos}}
\tablenotetext{c}{These limb-darkening parameters are fully described in \citet{Mandel} and obtained from 
\citet{MillerTwo}.}

\end{deluxetable}

\begin{figure}
\includegraphics[height = 3.5in, width = 2.25in, angle=270]{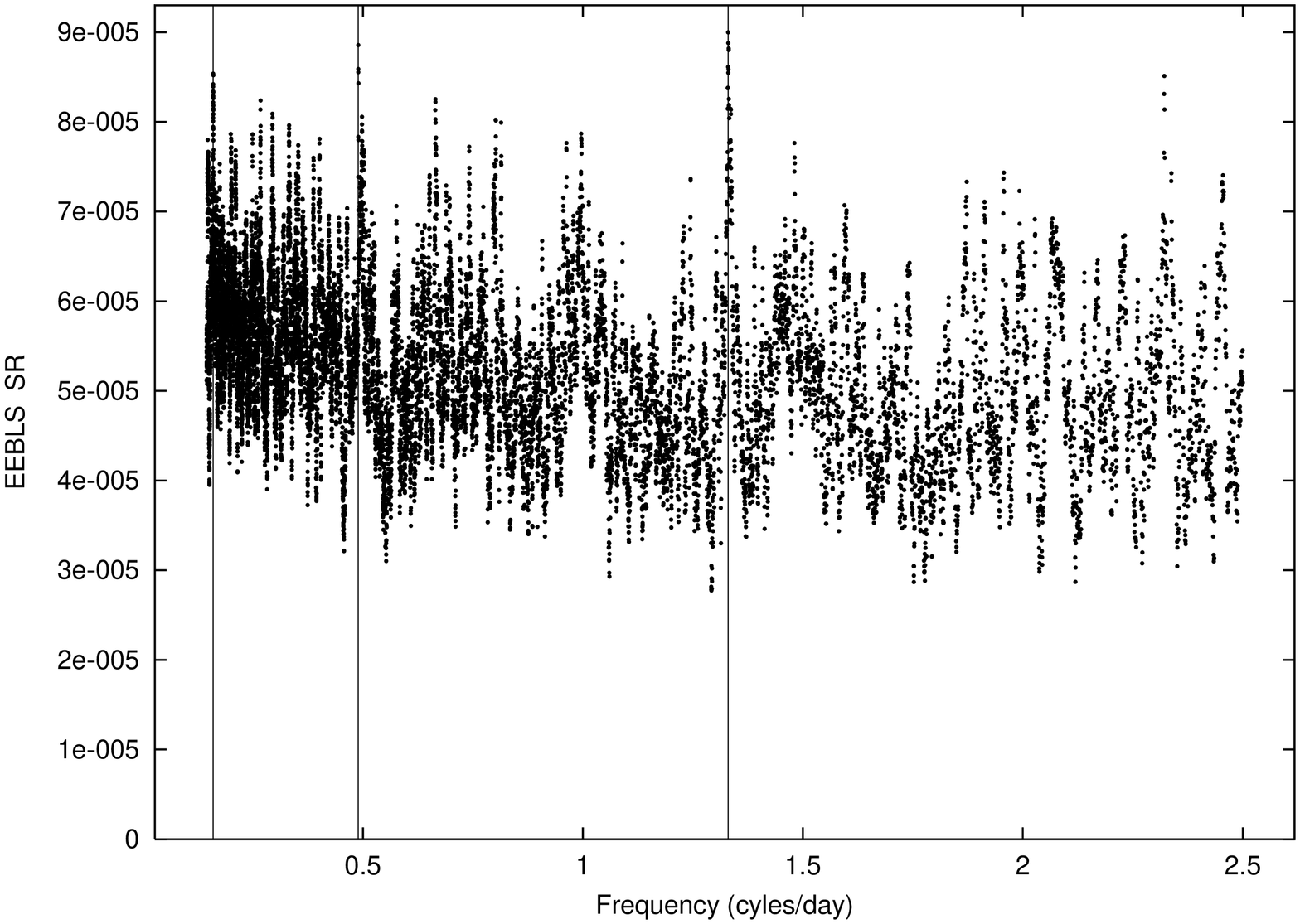} % 6, 4
\includegraphics[height = 3.5in, width = 2.25in, angle=270]{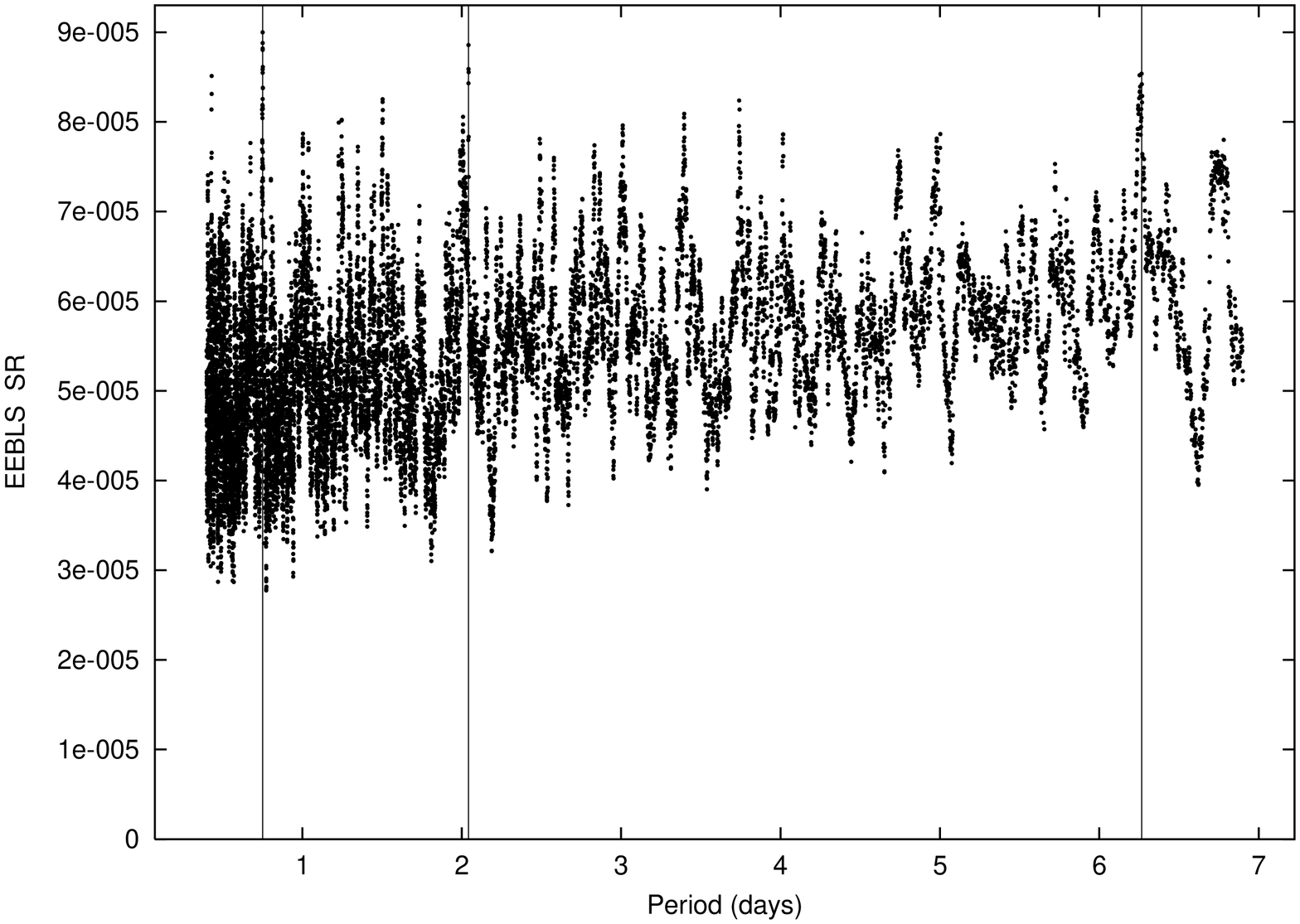}
\caption{{The EEBLS spectrum of the {\it MOST} HD 189733 data-set, plotted in frequency (top),
and period (bottom).
The \NumPassedSelection \ candidates that passed the transit selection
criteria are shown by the solid vertical lines.}  
\label{FigTransSelect}}
\end{figure}

\section{Monte Carlo statistics}
\label{SecMonte}

\subsection{Monte Carlo setup}

To assess the sensitivity of our routine and the {\it MOST} data-set
to smaller transiting planets in the HD 189733 system, simulated transits for planets of various radii
and orbital parameters were inserted 
into the {\it MOST} photometry and Monte Carlo statistics of the
transit recovery rate were generated. 
We thus proceed under the assumption that all signals in the data-set are noise,
and we place as sensitive of limits as possible under this paradigm. 
Realistic limb-darkened transits due to planets with various radii,
$R_{p}$, orbital phases, $\phi$, periods, $P$, and inclinations, $i$,
were inserted into the 2006 {\it MOST} HD 189733 data. The full non-linear model of \citet{Mandel} was used 
to ensure the inserted transits were as realistic as possible.
Circular
orbits were used. The M-dwarf companion star, HD 189733B, would have a negligible
effect on the orbit of a close-in putative planet in orbit around HD 189733, and thus the effect of this binary companion can be 
safely ignored. 

These
modified data were then subjected to the transit search algorithm
as discussed in $\S$\ref{SecTrans}. It should be noted that this includes
the automated filtering step,
and thus the synthetic transits were injected 
before the same filtering process our original data was subjected to,
therefore ensuring the validity of the following
Monte Carlo limits. 
Transits were inserted with logarithmic period spacing (as discussed
in \citealt{Croll}), with $\eta_{inp} = 0.04$ in the period
range $0.45 d < P_{inp} < 6.7 d$. In total, 36 period steps were
used for the 90$^{o}$ inclination angle cases.
For the 86$^{o}$ and 82$^{o}$ inclination angle cases transits were inserted with this same logarithmic
period spacing until the period exceeded 
the maximum period, or semi-major axis, that would produce a transit, a $<$ $(R_{*}+R_{P})$/cos$i$. 
For each trial period, simulated transits corresponding to 20
different exoplanet radii were inserted, sampling the period-radius
space of interest. For each period and radius at least \HowMany \ phases were
inserted.
For each of these points, the phase $\phi$ was generated
randomly to be in the range 0 $\le$ $\phi$ $<$ 1.
Because a 2.4 Ghz Pentium processor with 1 Gbyte of memory can perform
the transit search algorithm on an individual MOST HD 189733 data set
in $\sim$1 minute, exploration of the entire grid just mentioned for
all inclinations involves $\simeq$3$\times$$10^5$ iterations and 0.6 CPU years.
The calculation was performed on the LeVerrier Beowulf cluster in
the Department of Physics and Astronomy at the University of British
Columbia using 46 dual-CPU compute nodes.

An inserted transit
was judged to be detected if the parameters $\phi$ and $P$ returned
by the transit search algorithm were sufficiently close to the input
values: $\phi_{inp}$ and $P_{inp}$. The returned period had to satisfy
the following criteria: $|P - P_{inp}| < 0.05 d$ and $|P/P_{inp} - 1| <
1\%$. The limits of the criterion for $\phi$ were dependent on the planetary orbital period,
because as the period of the putative transiting planet decreases,
the fractional transit length increases
accordingly. Thus the accuracy required in the determination of
$\phi$ was relaxed for shorter periods. The criterion on $\phi$
is: $|\phi-\phi_{inp}|$ $<$ $0.09 - 0.0106$ $\times$ ($P_{inp}$ - 0.4$d$)/$d$.
This corresponds to accepting an error in phase of up to 9\% at 0.4$d$, but only up to
2\% at 7$d$.
Obvious multiples of the period of the inserted 
planet, up to $4$ times the inserted period, $P_{inp}$, 
as well as half-period ($P \approx \frac{1}{2}P_{inp}$), and one-third period ($P \approx \frac{1}{3}P_{inp}$)
solutions were also accepted. For the Monte Carlo tests when a transit was deemed recovered
the correct period was recovered 96\% of the time, 
while the remainder returned a period that was double, triple, quadruple, one-half, or one-third
of the inserted period, $P_{inp}$. 

The level above which a
transit recovered from the data-set
could be considered believable was also determined due to these Monte Carlo tests, at a level 
of $\Delta\chi^{2}\%$/$\Delta\chi^{2}_{-}\%$ $\ge$ $\TransAntiBelieve$ that
ruled out 95\% of the spurious candidates. Spurious candidates 
are defined as those candidates where the parameters returned by the 
transit search routine are different from the parameters inserted for the Monte Carlo tests. 
The motivation for this criterion is to ensure that a systematic event is
misidentified as a believable transiting planet less than
5\% of the time. Interestingly we can rigorously rule out these spurious candidates,
with our criterion set at a level where the most significant transit candidate
can be less significant than the most significant anti-transit (brightening) event.
Further discussion on this criterion is given in \citet{Croll}.
So as to not unduly skew this statistic we only generate this statistic using
putative planets with radii large enough that the routine has a realistic chance of detecting these
planets. We judge this to be planets greater than approximately the 25\% contour of our Monte Carlo statistics
without the RS criterion. Thus only planets with radii greater than the following limit are used: 
$R_{Pinp} > (0.10 + \frac{0.05}{6.3} \times (P_{in}-0.4d)/$d$ )$ $R_{J}$. 

\subsection{Monte Carlo results}

\begin{figure}
\epsscale{0.6}
\includegraphics[angle=270, scale = 0.52]{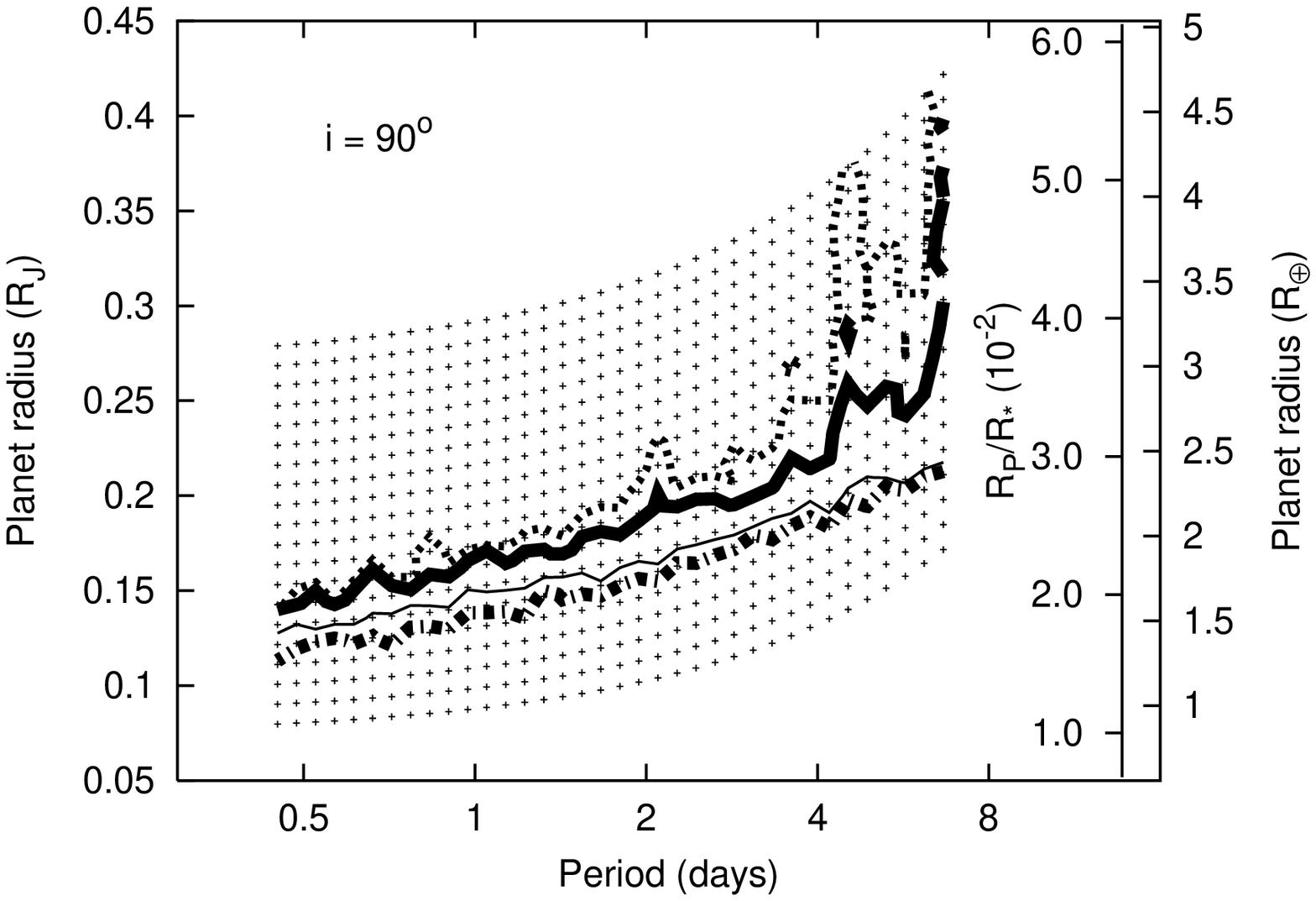}
\includegraphics[angle=270, scale = 0.52]{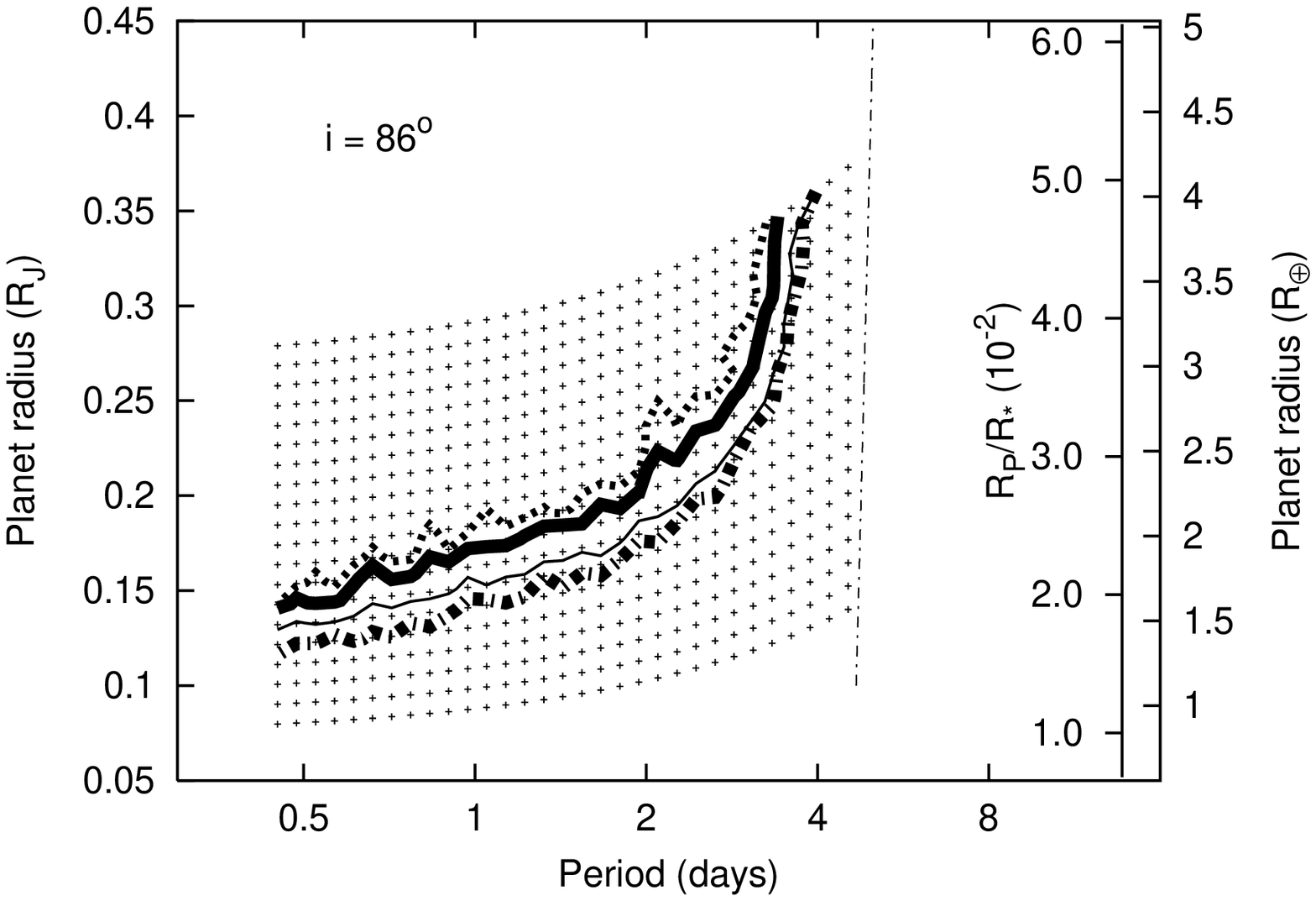}
\includegraphics[angle=270, scale = 0.52]{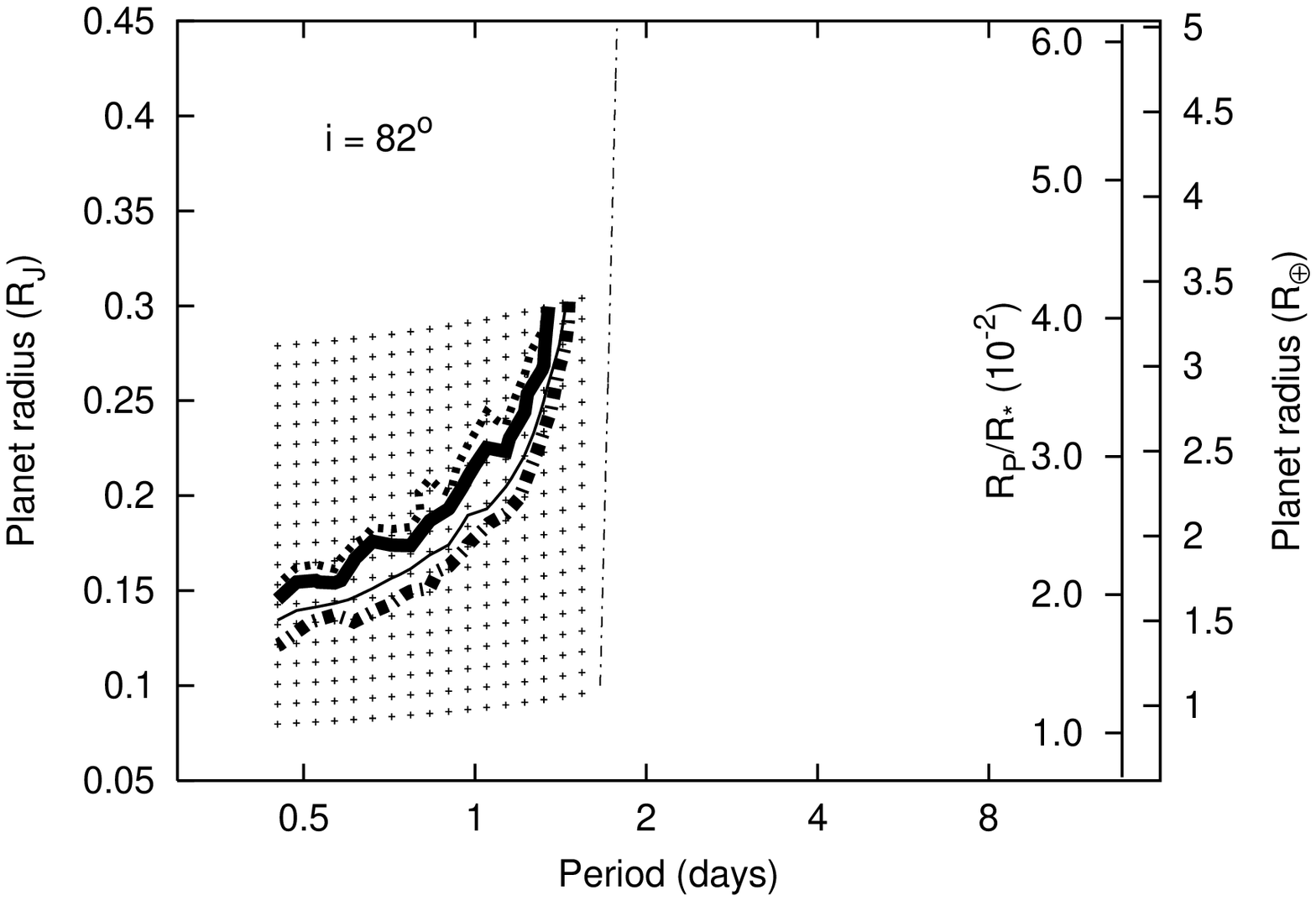}
\caption{
	{Confidence limits of transit detection as a function of planet radius and orbital period, for different orbital
	inclinations, based on Monte Carlo statistics. The crosses represent the radii and periods at which synthetic
	transits were inserted in the data. The dotted, thick-solid, and thin-solid lines represent the 99, 95 and 68\%
	confidence
	contours, respectively.
	The thick dot-dashed curve represents the 68\% confidence contour if the criterion $\Delta\chi^{2}\%$/$\Delta\chi^{2}_{-}\%$ $\ge$ \TransAntiBelieve \
	was not used.
	The near vertical dot-dash line in the later panels indicates 
	the maximum period that produces a transit for that given inclination angle.
	Note the logarithmic
	period scaling on the x-axis. 
	At least \HowMany \ phases were inserted for each of the radii-phase points.
	}  
	\label{FigMonte}
}
\end{figure}

The fractions of times that the artificially inserted transits were
recovered from the data for various radii, periods, and inclination
angles are given in Figure \ref{FigMonte}. Also indicated in Figure \ref{FigMonte} is the 68\% contour limit
that would be placed without the $\Delta\chi^{2}\%$/$\Delta\chi^{2}_{-}\%$ $\ge$ \TransAntiBelieve \ 
criterion.
The close agreement between the 68\% contour with or without this criterion
indicates that this criterion does not seriously impact the sensitivity of the Monte Carlo statistics generated, while providing
a robust limit so as to avoid false positives and spurious detections.

\begin{figure}
\epsscale{0.6}
\includegraphics[angle=270, scale = 0.52]{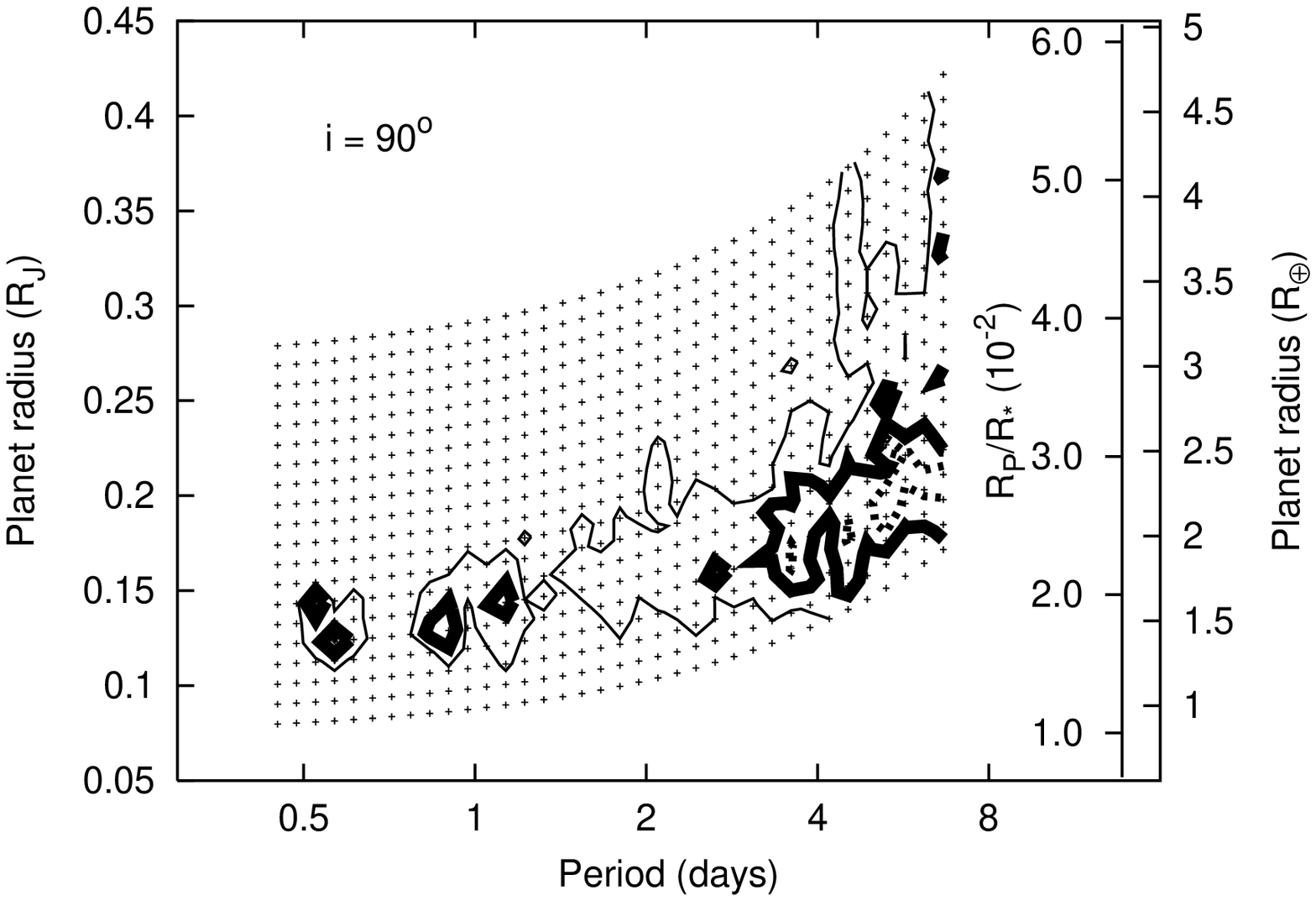}
\includegraphics[angle=270, scale = 0.52]{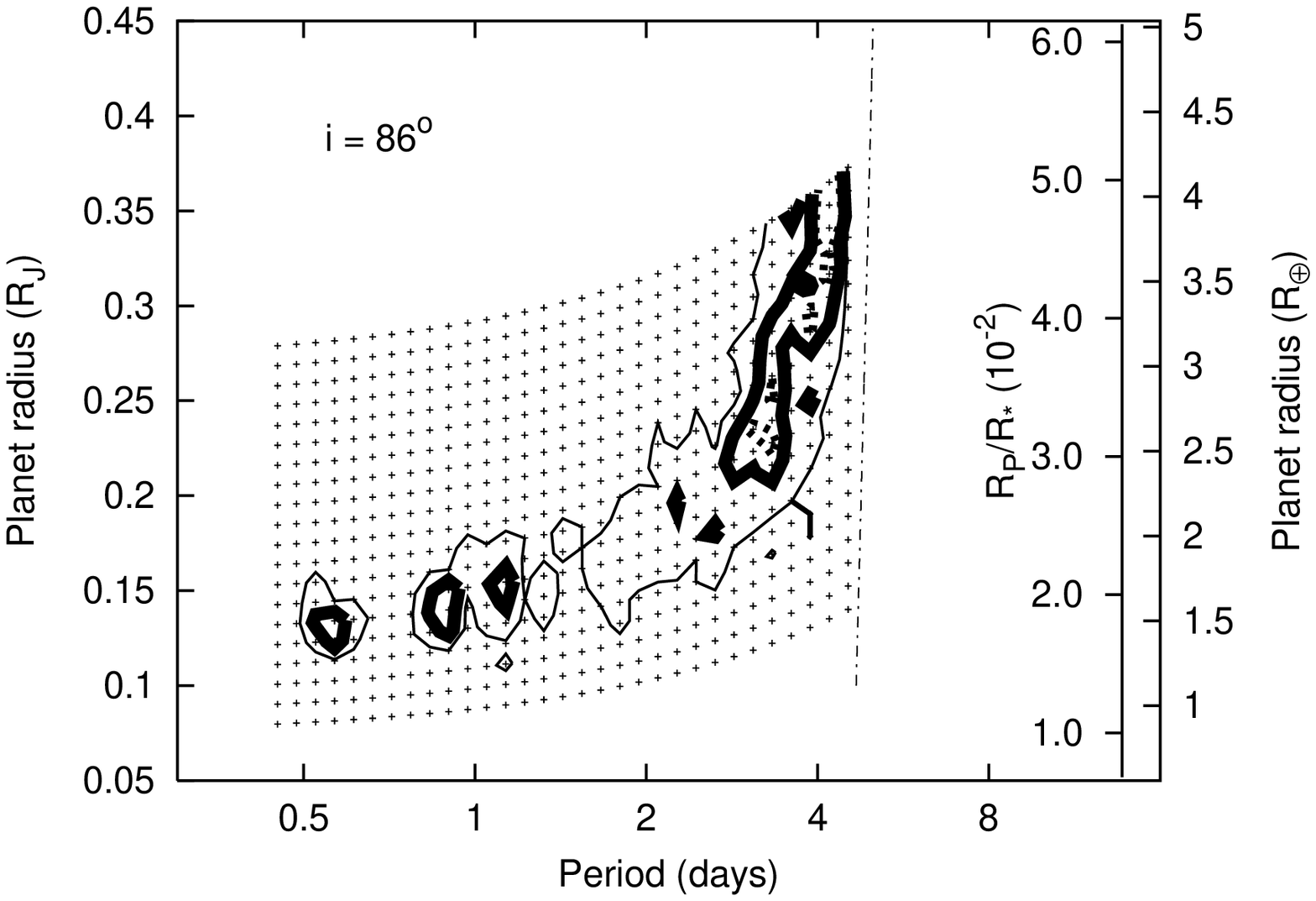}
\includegraphics[angle=270, scale = 0.52]{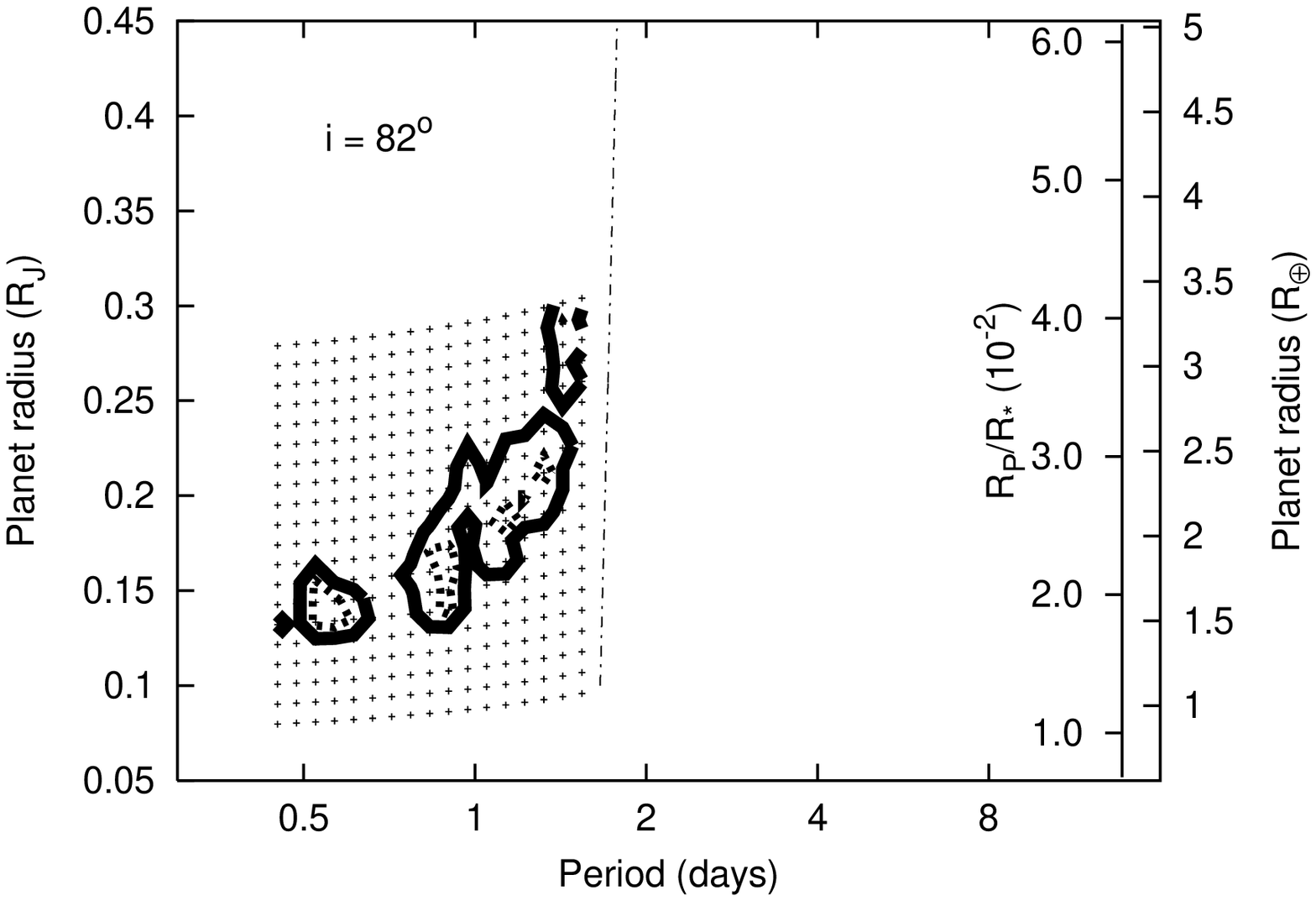}
\caption{
	{The likelihood of spurious transit detections returned by the Monte Carlo statistical analysis. The
	dotted, thick-solid and thin-solid lines represent the 10, 5 and 1\% spurious signal contours, respectively.
	The format of the figure is otherwise identical to Figure \ref{FigMonte}. Note that the 
	spurious signals occur in the intermediate regions of the period-radii space of interest,
	where it is not guaranteed that the correct transit will be recovered,
	but the inserted transit still causes significant deviations to the light-curve.
	}  
	\label{FigMonteFalse}
}
\end{figure}

We also explore the distribution of spurious signals across period-radii space.
These values are displayed in Figure \ref{FigMonteFalse}, and indicate that spurious signals
are negligible (below 1\%) other than the intermediate areas where the routine
has a small to moderate chance to correctly determine
the inserted transit.
This result is expected, as it is similar to the result of \citet{Croll} for HD 209458.

The performance of our routine is also expected to be slightly degraded 
at harmonics and subharmonics of the periods that the sinusoids 
were removed from the data at ($P$=1.0$d$, and $P$=0.5$d$). 
Due to the fact that 
the fractional length of a transit is comparitively large for short periods of 0.5 - 1.0 $d$,
it can be difficult to
completely differentiate and properly recover a transit signal near the
harmonics and subharmonics of these periods in all cases.
The period ranges demonstrating these drops in survey sensitivity
are very narrow ($\frac{ \Delta P }{ P }$ $\approx$ 2\%).
Limited statistical tests confirm that the loss in sensitivity near these periods is quite small,
similar to that of HD 209458 \citep{Croll}.

The harmonics
and subharmonics of the orbital period of the known planet HD 189733b, $P$ $\approx$ 
2.218573d \citep{Bakos},
are also expected to show decreased sensitivity to transits 
due to the fact that the transits of the known planet were excised from the data. 
This reduction of sensitivity
at these periods is unfortunate considering that it is expected that low-mass planets show a preference for 
these mean-motion resonant orbits (\citealt{Thommes}; \citealt{Papaloizou} and \citealt{Zhou}). 
The fractional length in phase that was removed at $P$ $\approx$ $2.218573$ to exclude the transit of the
known planet was 4.0\%
of the total.
Thus for the subharmonics of the known planet ($P$ $\approx$ 4.44$d$, 6.66$d$) in 
approximately 4.0\% of the
cases the routine will be unable to recover a putative transiting planet, as the
transits will coincide with the transits of the known planet, and this information 
will have been completely removed.
For
the harmonics of the orbital period of HD 189733b, the
situation is better, since not all of the transit events
would have been removed from the data.
For periods near the first harmonic ($P$ $\approx$ 1.11$d$), every second occurance of a transit will be lost
in only 8.0\% of the cases. For the second harmonic ($P$ $\approx$ 0.74$d$),
every third transit would be missed in 12.0\% of the cases.
For periods near these values the sensitivity limits can be expected to be 
only slightly degraded from those shown in Figure \ref{FigMonte}. Even for exact harmonics 
the routine's sensitivity should be only marginally worse than the limits quoted in Figure \ref{FigMonte},
as the remaining transits should allow the correct period (or a multiple of the period) to be recovered.
The mean-motion resonances from the known exoplanet HD 189733b have been examined 
using the transit-timing technique
with ground-based photometry by \citet{WinnPsi}, HST photometry by \citet{PontPrep},
and with {\it MOST} photometry 
by \citet{MillerTwo}.
It should also be noted that the detection limits presented
here are valid for circular orbits, but are largely applicable to orbits of other eccentricities,
as summarized for the analogous case of HD 209458 in \citet{Croll}.

We have made the simplifying assumption that the hypothetical planet
that transits the star to produce the synthetic transits of our Monte Carlo statistics does
not pass over prominent starspots along our line of sight.
The known planet has not been found to 
occult a prominent starspot in the {\it MOST} photometry \citep{MillerTwo},
although a preliminary investigation of HST photometry
of HD 189733 indicates that the planet does occult small
to moderate sized starspots during their observations
\citep{PontPrep}.
Given that the rotational
period of the star is greater
than the period
range that is investigated here, 
the limits presented here would likely only be slightly adversely affected 
if this hypothetical planet does pass in front of a prominent starspot. This is because
the impact on the transit dip of the planet occulting
a large starspot would be relatively short, and should thus not severely affect our limits.  

Thus the search routine described above in $\S$\ref{SecTrans}
should be able to detect planets with radii greater than the limits
given in Figure \ref{FigMonte} with the {\it MOST} data. In the most optimistic case of an edge-on $90^{o}$ inclination angle transit, for planets
with periods from approximately half-a-day to one week, this limit is
approximately \Rjupsmall \ to \Rjuplarge \ $R_{J}$, or \Rearthsmall \
to \Rearthlarge \ $R_{\earth}$, respectively, with 95\% confidence. 
These limits would scale up, or down linearly if the expected stellar radius, 
$R_{*}$ = 0.755 $R_{\odot}$, is an over- or under-estimate, and are thus perhaps better
expressed as $R_p$/$R_{*}$ = \Rratiosmall \ and $R_p$/$R_{*}$ = \Rratiolarge \ for periods
from half-a-day to one week, respectively.  
For periods at the upper range of those investigated here ($P$ $>$ 6.5$d$) 
the routine is unable to pick out
the transit in a low percentage of cases, because the cubic spline
has removed significant power at that period; this effect can be observed
as a reduction of sensitivity for longer periods 
in the ninety-degree inclination angle ($i$=90$^{o}$) panel of Fig. \ref{FigMonte}.
The performance of our automatic routine in this period-radius space probably underestimates
the true sensitivity of our survey, as it would be obvious that the cubic spline
has overcorrected in these cases through an individual inspection of the phased curve.
Also, limited statistical tests have shown that marginally improved limits could
probably be set for the shortest periods (P $\le$ 3$d$) by excluding the last seven days
of observations with increased noise, and only producing these limits with the first 14 days
of the highest quality data.  

%  If one assumes a mean density of $\rho$ $\approx$ 3000 kg m$^{-3}$ 
% - a value averaging the various extrasolar Super-Earth models discussed below -  
% their respective masses would be \Massjupitersmall \ and \Massjupiterlarge \ $M_{J}$,
% or \Massearthsmall \ and \Massearthlarge \ $M_{\earth}$, respectively.
% If the putative planet were gaseous, and thus a possible hot Neptune analogue ($\rho$ $\approx$ 1600 kg m$^{-3}$),
% our upper radius limit, $R_{P}$ = \Rearthlarge \ $R_{\earth}$, would result
% in a planetary mass of \MassNeptune \ $M_{\earth}$.
% Thus the mass-period parameter space
% we have ruled out in this study, assuming this hypothetical mean density,
% is the most sensitive limit set to date in this system for near edge-on inclination angles for our period range.

\section{HD 189733 transit search}
\label{SecHDTrans}

The {\it MOST} HD 189733 2006 data-set was submitted to the analysis outlined in $\S$\ref{SecTrans}.
There were no transiting planet candidates that met the detection criteria. 
As the Monte Carlo tests indicate we would be able to detect 
exoplanets in this system with periods 
from half-a-day to one week with radii larger than the radius limits given in Figure
\ref{FigMonte}, we can safely rule out these sized-exoplanets in this system. 
 
\begin{figure}
\includegraphics[width=1.5in, height = 3.5in, angle=270]{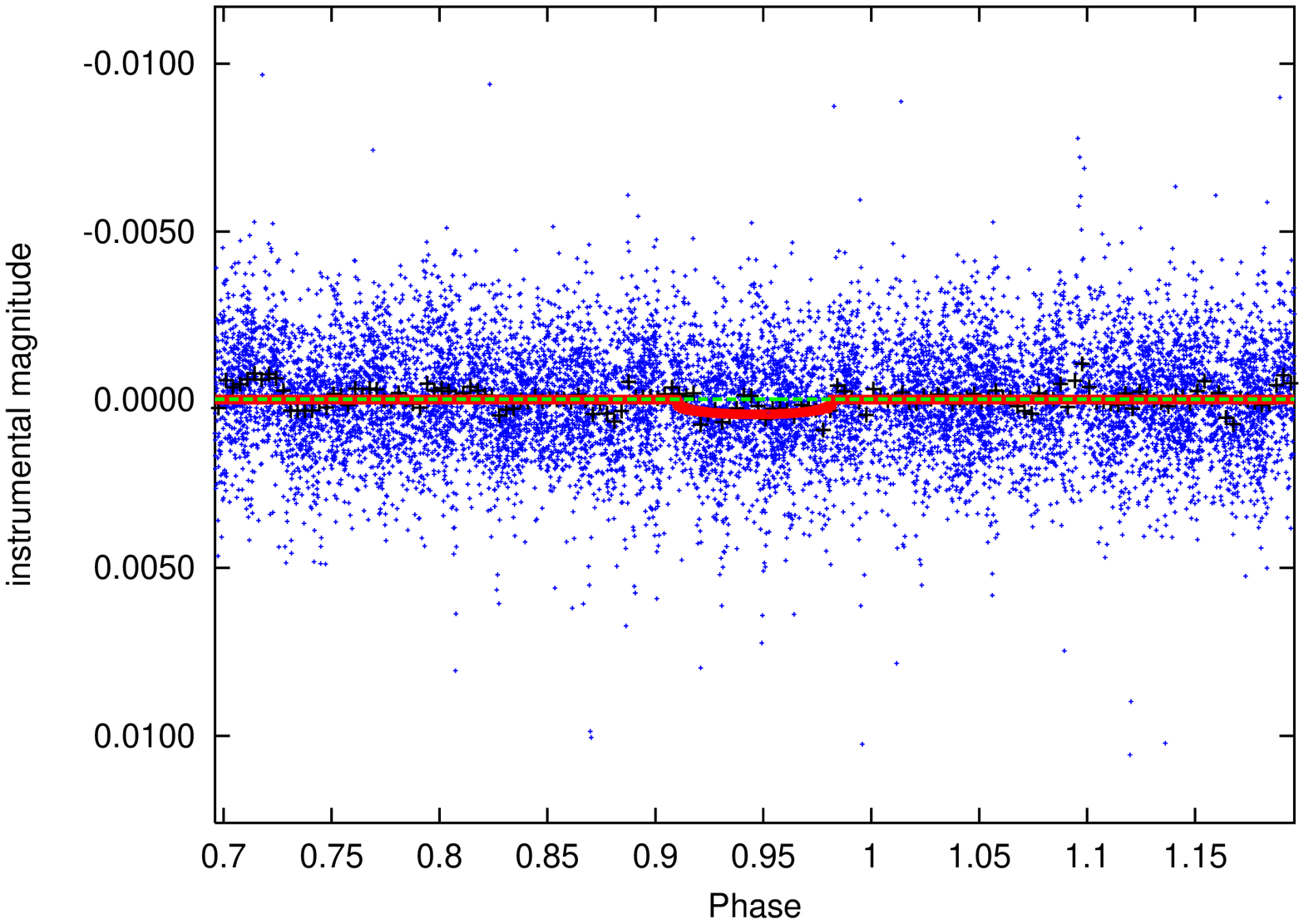} % 2.3, 6.0
\includegraphics[width=1.5in, height = 3.5in, angle=270]{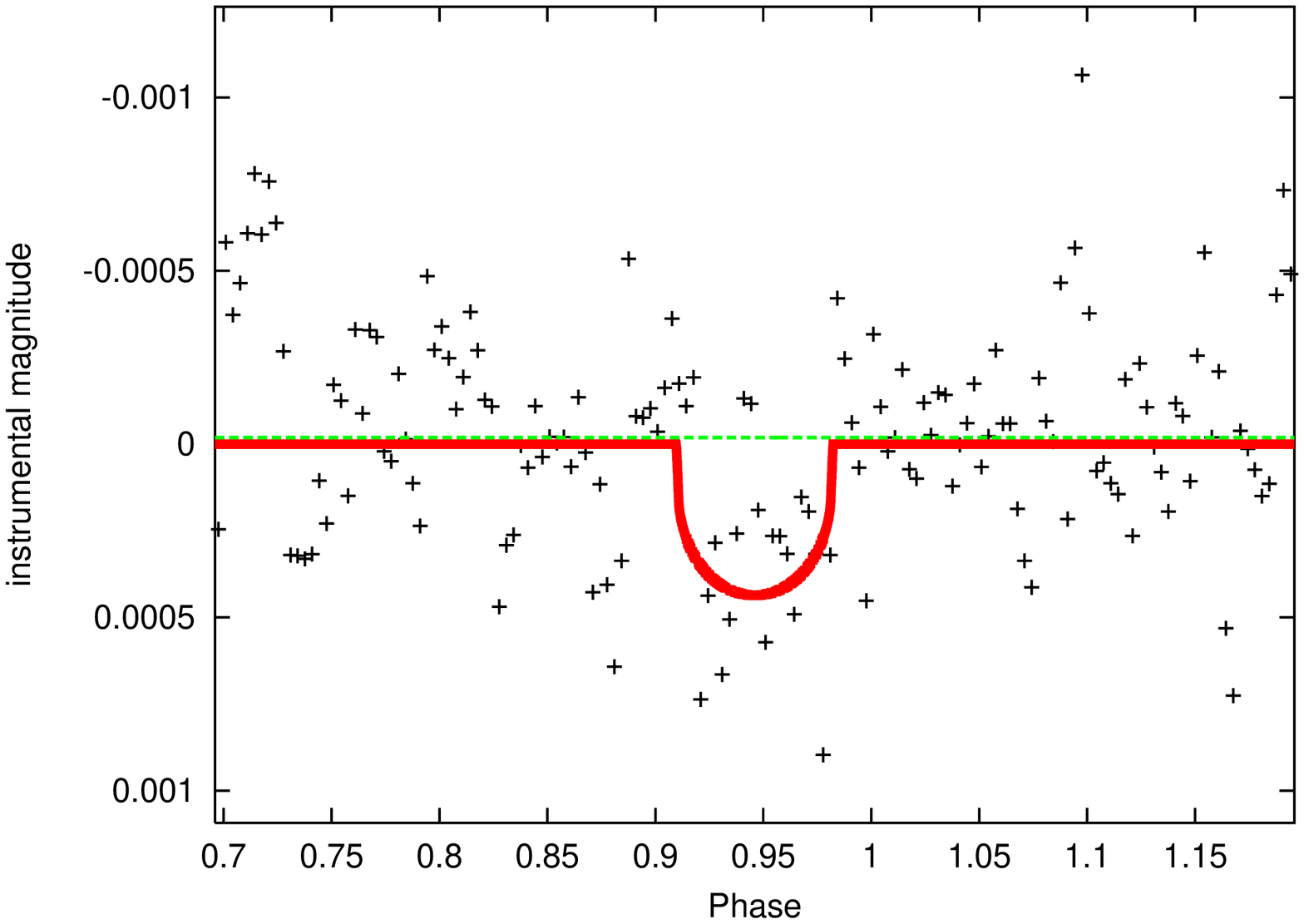}
\caption{
	{The best transit candidate as identified by our analysis of {\it MOST}'s HD 189733 2006
	data-set, but with marginal significance. The thick red-line represents the transit model,
	while the thin green-dashed line represents the constant
	brightness model.
	At top, the unbinned data are shown in blue, and the binned data are shown in black;
	at bottom the binned data only.
	As this candidate failed the improvement in transit over anti-transit ratio statistic (RS) criterion,
	we report it as a possible, but unlikely candidate. 
	The period and radius of this putative planet would be
	approximately \PlanetPeriod$d$ and \PlanetRadius $R_{J}$ (\PlanetRadiusEarth$R_{\oplus}$).}  
	\label{FigCandidate}
}

\end{figure}

We also present the 
details of the candidate with the greatest improvement in $\chi^{2}$, even though evidence for its
existence is marginal. 
The phased diagram of this candidate (Figure \ref{FigCandidate}) is not particularly convincing,
and 
this candidate did not meet the RS detection criterion, 
($\Delta\chi^{2}\%$/$\Delta\chi^{2}_{-}\%$ $\approx$ \PlanetChiPerc \ $\not\ge$ $\TransAntiBelieve$).
For these reasons we do not report it as a probable transiting candidate. 
The specifics of this putative candidate are
given in Table \ref{TableCandidate}.
This event, with a period of $P \approx \PlanetPeriod$$d$, 
has modest statistical significance, and thus 
we report it is a possible, but unlikely, candidate. This candidate 
would correspond to a putative hot Super-Earth, with
period P = \PlanetPeriod $d$, radius $R_{p}$ = \PlanetRadius $R_{J}$ (\PlanetRadiusEarth $R_{\oplus}$), 
and inclination $i$ = \PlanetInc$^{o}$. Interestingly the period of this planet places it quite close to the
1:3 interior mean-motion resonance to the known planet ($P$$\sim$0.74$d$).
For $\rho$ $\approx$ 3000 kg m$^{-3}$ the mass of this putative planet would be
approximately \PlanetMassEarth \ $M_{\earth}$ (\PlanetMassJupiter \ $M_{J}$).
A transit time combined with the supposed period is given in Table \ref{TableCandidate}. 
Evidence for a transit at this period is marginal, but additional {\it MOST} photometry of the HD 189733
system should confirm or deny its existence.
% The transit timing investigation of HD 189733 of
% \citet{MillerTwo} should also be able to rule out a possible candidate near this mean-motion resonance.

The most significant brightening event was one observed with a period of approximately $P \approx 6.28 d$.
It is likely statistical in nature, and thus is not expected to be related to any specific astrophysical process.

Although our transit search covers a continuous set of periods, we do
not expect the existence of another planet with a semimajor axis `too close'
to that of the known planet HD189733b, since the gravitational perturbations
of the known Jupiter-scale planet will destabilize the orbit of another
object.
Taking the known planet's data - $a_b$=0.0313~AU, and mass
$M_b$=1.15~$M_{Jup}$ \citep{Bouchy} in orbit around a star of mass 
$M_*$=0.82 solar masses \citep{Bakos} - orbits in an
annulus
$ a_{in} = (1-\Delta) a_b < a_b < (1+\Delta) a_b = a_{out} $
will be unstable. The fractional width $\Delta$ of the
annulus is given by \citep{Gladman}:
$ \Delta = 2.40 ( M_b / M_* )^{1/3} $
where the assumption is being made that the mass of the second
planet is much less than the known planet (if not, the width
of the zone scales as the cube root of the sum of the masses of the planets).
Here $\Delta$=0.26, and so we calculate that planets with periods in the
range 1.41 -- 3.14 days will be intrinsically unstable due to the
presence of the known planet.
Thus, some of the parameter space searched in Figure \ref{FigMonte} was very
unlikely to have been occupied by a second planet - Trojans lagging or leading \citep{Ford} HD 189733b being a
notable exception.

We can also exclude transits from planets with longer orbital periods. There were no 
transits from edge-on ($i$ = $90^{o}$) planets with radii greater than $\sim$0.35 
$R_{J}$ (3.9 $R_{\earth}$) during {\it MOST}'s observations of HD 189733,
as a planet of this size would be readily visible with even a cursory 
inspection of the data. A planet of this size would correspond to $\sim$ 20 
$M_{\earth}$, as calculated from the 
models of \citet{Fortney} assuming a composition similar to Neptune. We can 
also place a limit on transiting planets with orbital periods from 7 $d$ $<$ P $<$ 
10.5 $d$ - in this region we would expect two transits in our data-set. Limited 
statistical tests indicate that we can rule out planets with radii greater than 
approximately $\sim$0.32 $R_{J}$ (3.6 $R_{\earth}$) in this region - this 
corresponds to $\sim$14 $M_{\earth}$ \citep{Fortney}, again assuming a 
composition similar to Neptune. We do not provide a formal limit from Monte 
Carlo estimates for this range of periods as we prefer to restrict our formal transit 
search to periods that will result in three transits in our data-set.

As an additional sanity check this transit search method was applied to the current data-set
without the removal of the transits
of the known planet, HD 189733b. 
As the transits are apparent in the unbinned data (Figure \ref{FigData}), this causes
the cubic-spline to over-correct and remove a moderate portion
of the signal at this period. Even so, the routine correctly uncovers the transit of HD 189733b with reasonable accuracy.
The actual parameters of HD 189733b \citep{Bakos} were recovered within
0.04\% and 0.3\% in period and phase, respectively, while the inclination angle
and radius were recovered within 1.0$^{o}$ and 0.20$R_{J}$ of the actual parameters. 
The moderate discrepancy between
the inserted and recovered radius is due to the fact that the cubic spline overcorrects and
removes a portion of the transit-signal.
  
%%%%%%%%%%%%
% Actual Period P=2.218573
% Eph 2453629.39420 -2451545 = 2084.3942 + 144*P = 2403.868712
% i = 86
% R = 1.154 (all of these from Bakos, Knutson 2006)
%%%%%%%%%%%%

\begin{deluxetable}{cc}
\tabletypesize{\footnotesize}
\tablecaption{Transit candidate of marginal significance \label{TableCandidate}}
\tablewidth{0pt}
\tablehead{\colhead{parameter}&		\colhead{value} \\}
\startdata

$P$								& \PlanetPeriodLong d \\
$R_{p}$ 							& \PlanetRadiusEarth $R_{\oplus}$ (\PlanetRadius $R_{J}$) \\ 
$i$								& \PlanetInc$^{o}$ \\
Ephemeris Minimum (JD-2451545)					& \PlanetEphemeris \\
$\Delta\chi^{2}\%$/$\Delta\chi^{2}_{-}\%$ 			& \PlanetChiPerc \\
Mass (assuming $\rho$ $\approx$ 3000 kg m$^{-3}$)		& \PlanetMassEarth \ $M_{\earth}$ (\PlanetMassJupiter \ $M_{J}$)\\
\enddata
\end{deluxetable}

\section{Discussion}
\label{SecDiscuss}

{\it MOST}'s 2006 observations of HD 189733
have been searched for evidence of other exoplanets in the system. 
The transit-search routine described in \citet{Croll} has been used to search for
realistic limb-darkened transits to take advantage of the precise, near-continuous photometry
returned by {\it MOST} and the fact that the 
stellar characteristics of the star, HD 189733, are well established.
Monte Carlo statistics were generated using this routine and indicate that
this routine in combination with the aforementioned {\it MOST} data have placed robust
limits in the size of transiting bodies that have been ruled out in this system 
for a range of periods and inclination angles. In the most optimistic case 
of edge-on transits, planets with radii greater than
\Rjupsmall \ $R_{J}$ ($\sim$\Rearthsmall \ $R_{\oplus}$) to \Rjuplarge \ $R_{J}$ ($\sim$\Rearthlarge \ $R_{\oplus}$) with
periods from half-a-day to seven days, respectively,
have been ruled out with 95\% confidence through this analysis. For orbits co-planar with
that of the known planet ($i$ $\approx$ 86$^{o}$)
the limits are similar, except for longer periods ($P$$>$4$d$). For these longer periods the planets cease
to transit
the star along our line of sight and thus we cannot place firm limits for periods greater than 4-days at
$i$ $\sim$ 86$^{o}$, or for $i$ $<$ 86$^{o}$.
Specifically, we have been able to rule out
transiting planets in this system with radii greater
than those given in Figure \ref{FigMonte}. 
This work has constrained theories that predict hot Super-Earths, and hot Neptunes
due to the inward migration of 
HD 189733b, such as those proposed by \citet{Zhou}, \citet{Raymond}, \citet{Mandell}, 
and \citet{Fogg}.
% Further {\it MOST} photometry should be able to improve the
% radii-period limit that was set in this work, and further constrain or confirm these thories that predict
% hot Earths, and hot Neptunes in nearby orbits to the 
% known planet.

It should also be noted that we are able to place very sensitive limits on the size of planets that can be ruled
out in this system despite the prominent $\sim$3\% rotational modulation observed in this system. By removing
the observed rotational modulation using a cubic spine with a binning time longer than the expected duration
of the transit of the planet across the star, we are able to remove the effects of the prominent modulation
while remaining sensitive to the transits
of planets as small as those expected from hot Super-Earths and hot Neptunes. These methods should
be applicable to current or future space-based transit search missions, such as COROT and Kepler.

 We are also able to place limits on the size of Trojan planets that have been ruled out in this analysis, 
assuming the Trojan consistently transits the star with an inclination angle close to that
of the known planet ($i$$\approx$86$^{o}$).
We use the definition of Trojans given by \citet{Ford} as objects occupying the L4 or L5 stable Lagrange points.
Trojans lagging or leading
HD 189733b with a radius above $R$ = \RTrojanJupiter \ $R_{J}$ (\RTrojanEarth $R_{\oplus}$) should have
been detected with 95\%
confidence. Using a mean density of $\rho$$\approx$ 3000 kg m$^{-3}$ this corresponds
to \MTrojanEarth \ $M_{\oplus}$.
A detailed study of this photometry should be able to radically improve this limit on the size of Trojans that can be ruled out with
this photometry, assuming the Trojan consistently transits the star along our line of sight.

\subsection{Mass constraints on other Exoplanets in the system}

The above limits on the smallest planetary radii excluded by {\it MOST}
for transiting planets allow some estimates on the type of planets
that are being excluded in this system.
For Gas Giant planets, including hot Neptunes, radii change with time as the planet
cools, so we make use of the fact that the age of the HD 189733 system
is greater than 1 Gyr \citep{Melo}. We follow the suggestion of \citet{WinnPsi}
and adopt an age of $\sim$4.5 Gyr.
We thus can refer to cooling
models similar to the planets in the Solar System. One exception would
be the consideration of possible planets in orbits smaller than that of
HD189733b, where tidal heating will lead to larger radii per given mass
and composition.

Our radius limit of \Rearthsmall \ - \Rearthlarge \ $R_{\earth}$ already places us in the realm of Super-Earth
and hot Neptune planets. \citet{Fortney} has recently provided theoretical predictions for exoplanetary
radii for a diverse range of possible exoplanetary compositions, including hot Super-Earth, and hot Neptune planets.
The Super-Earth models of \citet{Fortney} agree well with the more detailed Super-Earth
models of \citeauthor{ValenciaA} (\citeyear{ValenciaA}, \citeyear{ValenciaB}).
For our Super-Earth exoplanetary limits we use equation 7 and 8 of \citet{Fortney}. 
Our inner period limit of $R_{P}$ $<$ \Rearthsmall $R_{\earth}$ allows us to rule out
planets more massive than \MearthsmallIce $M_{\earth}$ for a pure ice planet, \MearthsmallRock $M_{\earth}$ for a pure rock planet,
and \MearthsmallIron $M_{\earth}$ for a pure iron planet. 
Our outer period limit of $R_{P}$ $<$ \Rearthlarge $R_{\earth}$ only allows to place a useful limit on pure ice Super-Earth planets.
At this limit we are able to rule out planets more massive
than \MearthlargeIce $M_{\earth}$ for pure ice planets. Pure rock and pure iron planets of this size are moderately less massive,
and comparable
in mass, respectively, to HD 189733b itself - planets of this mass, of course,
have already been ruled out in this system with radial-velocity
measurements (\citealt{WinnLambda} as quoted in \citealt{MillerTwo}).
We note that our limit on ice planets may be more sensitive than the limits quoted
here as one would expect that ice planets at such small orbital periods would have steam atmospheres, and would thus 
be significantly inflated at these small orbital seperations \citep{Kuchner,Fortney}.

We are also able to place limits on the mass of hot Neptune-type planets that have been ruled out in our analysis.
Hot Neptune planets should have cores similar in density to giant ice-planets, and an outer atmosphere of H/He \citep{Fortney} - they
should thus be moderately less dense than giant-terrestrial ice planets, and thus be of a similar density to Neptune in our own solar
system \citep{Fortney}.
Our inner period radius limit ($R$ $<$ \Rearthsmall $R_{\earth}$) excludes hot Neptunes entirely, as
hot Neptune exoplanets are expected to be of similar size to Neptune (3.9 $R_{\earth}$)
(\citealt{Fortney}; \citealt{ValenciaA}; \citealt{ValenciaB}).
At our outer period radius limit ($R$ $<$ \Rearthlarge $R_{\earth}$) we can rule out hot Neptune
exoplanets more massive than
\MassNeptune \ $M_{\earth}$. 

The radius limits presented here,
combined with the mass-limits from transit-timing analyses (\citealt{WinnPsi}; \citealt{MillerTwo}; \citealt{PontPrep}),
place very firm constraints
on the sizes and masses of bodies that could still reside in close orbits to HD 189733.
Re-observations of this system planned with {\it MOST}
should be able to further reduce the size of planets that could remain undetected in this
system, and further constrain theories that predict hot Super-Earth, and hot Neptune planets in similar orbits to
that of the known hot Jupiter exoplanet HD 189733b. HD 189733 thus joins the growing list of hot Jupiter exoplanetary systems 
(HD 209458 [\citealt{Croll,Miller,Agol}], TrES-1 [\citealt{Steffen}]) 
for which firm constraints have been placed 
on the size of other exoplanets in similar orbits to that of the known hot Jupiter by a variety of methods.

\acknowledgements
We would like to thank the anonymous referee for a constructive review.
The Natual Sciences and Engineering Research Council of Canada supports the research of D.B.G., J.M.M., A.F.J.M., J.F.R., S.M.R., G.A.H.W., and B.G..
Additional support for A.F.J.M. comes from FCAR
(Qu\'ebec). R.K. and A.W., are supported by the Canadian Space Agency. W.W.W. is supported by the Austrian Space Agency and the Austrian Science Fund (P17580).
The Canadian Foundation for Innovation, and the BC Knowledge Development Fund
provided funding for the
LeVerrier Beowulf cluster.


\begin{thebibliography}{}

\bibitem[Agol \& Steffen (2007)]{Agol}
Agol, E. and Steffen, J.~H. 2007, \mnras, 374, 941

\bibitem[Baglin (2003)]{Baglin}
Baglin, A. 2003, Advances in Space Research, Volume 31, 345-349.

\bibitem[Baines et al.(2007)]{Baines}
Baines, E.~K. et al. 2007, astro-ph/0704.3722

\bibitem[Bakos et al.(2006a)]{Bakos}
Bakos, G.A., Knutson, H., Pont, F., Moutou, C., Charbonneau, D., Shporer, A., Bouchy, F., Everett, M., Hergenrother, C., Latham, D.W., Mayor, M., Mazeh, T., Noyes, R.W., Queloz, D., Pal, A., Udry, S. 2006, \apj, 650, 1160

\bibitem[Bakos et al.(2006b)]{BakosPal}
Bakos, G.A., Pal, A., Latham, D.W., Noyes, R.W., Stefanik, R.P. 2006. \apjl, 641, L57

\bibitem[Barge et al.(2005)]{Barge}
Barge, P., Baglin, A., Auvergne, M., Buey, J.-T., Catala, C., Michel, E., Weiss, W. W., Deleuil, M., Jorda, L., Moutou, C.,
COROT Team 2005, SF2A-2005: Semaine de l'Astrophysique Francaise, eds. F. Casoli, T. Contini, J.M. Hameury \& L. Pagani, EdP-Sciences, Conference Series, p. 193.

% Kepler Paper Review Article Good
\bibitem[Basri et al.(2005)]{Basri}
Basri, G., Borucki, W.~J., Koch, D. 2005, New Astronomy Review, 49, 478.

\bibitem[Borucki et al.(2004)]{Borucki}
Borucki, W.; Koch, D.; Boss, A.; Dunham, E.; Dupree, A.; Geary, J.;
Gilliland, R.; Howell, S.; Jenkins, J.; Kondo, Y.; Latham, D.; Lissauer,
J.; Reitsema, H. 2004, Second Eddington Workshop: Stellar structure and
habitable planet finding, eds. F. Favata, S. Aigrain \& A. Wilson,
ESA SP-538, Noordwijk: ESA Publications Division, 177 - 182.

\bibitem[Bouchy et al.(2005)]{Bouchy}
Bouchy, F., Udry, S., Mayor, M., Moutou, C., Pont, F., Iribarne, N., Da Silva, R., Ilovaisky, S., Queloz, D., Santos, N.C., Segransan, D., Zucker, S. 2005, \aap, 444, L15

\bibitem[Burke et al.(2006)]{Burke}
Burke, C.J., Gaudi, B.S., DePoy, D.L., Pogge, R.W. 2006, \aj, 132, 210

\bibitem[Burke et al.(2007)]{BurkeSeven}
Burke, C.J. et al. 2007, \apj, astro-ph/0705.0003

\bibitem[Collier-Cameron et al.(2006)]{Collier}
Collier-Cameron et al. 2006, \mnras, 373, 799

\bibitem[Croll et al.(2007)]{Croll}
Croll, B., Matthews, J.M., Rowe, J.F., Kuschnig, R., Walker, A., Gladman, B.,
Sasselov, D., Cameron, C., Walker, G.A.H., Lin, D.N.C., Guenther, D.B., Moffat, A.F.J.,
Rucinski, S.M., Weiss, W.W. 2007, \apj, 658, 1328

\bibitem[Croll et al.(submitted)]{CrollSpot}
Croll, B. et al. \apj, submitted 13 Apr 2007 

\bibitem[Deeg et al.(1998)]{Deeg}
Deeg, H.~J. et al. 1998, \aap, 338, 479


\bibitem[Deming et al.(2006)]{Deming}
Deming, D., Harrington, J., Seager, S., Richardson, L.~J. 2006, \apj, 644, 560

\bibitem[Doyle et al.(2000)]{Doyle}
Doyle, L.~R. et al. 2000, \apj, 535, 338

\bibitem[Fogg \& Nelson (2007)]{Fogg}
Fogg, M.~J., Nelson, R.~P. 2007, \aap, 461, 1195

\bibitem[Ford \& Gaudi (2006)]{Ford}
Ford, E., Gaudi, B.S. 2006, \apjl, 652, L137

\bibitem[Fortney (2007)]{Fortney}
Fortney, J.~J., Marley, M.~S., Barnes, J.~W., 2007, \apj, 659, 1661

\bibitem[Gaudi \& Winn(2007)]{Gaudi}
Gaudi, B.~S., Winn, J.~N. 2007, \apj, 655, 550

\bibitem[Gladman(1993)]{Gladman}
Gladman, B. 1993, Icarus, 106, 247

\bibitem[Grillmair et al.(2007)]{Grillmair}
Grillmair, C.~J., Charbonneau, D., Burrows, A., Armus, L., Stauffer, J., Meadows, V., Van Cleve, J., Levine, D.,
2007, \apjl, 658, L115

\bibitem[Hebb et al.(2007)]{Hebb}
Hebb, L. et al. 2007, astro-ph/0704.3584

\bibitem[Kov{\'a}cs et al.(2002)]{Kovacs}
Kov{\'a}cs, G., Zucker, S., Mazeh, T. 2002, \aap, 391, 369

\bibitem[Knutson et al.(2007)]{Knutson}
Knutson, H.A. et al. 2007, Nature, 447, 183

\bibitem[Kuchner (2003)]{Kuchner}
Kuchner, M.J. 2003, \apjl, 596, L105

\bibitem[Mandel \& Agol(2002)]{Mandel}
Mandel, K., Agol, E. 2002, \apj, 580, L171

\bibitem[Mandell et al.(2007)]{Mandell}
Mandell, A.~M., Raymond, S.~N., Sigurdsson, S. 2007, astro-ph/0701048

\bibitem[Matthews et al.(2004)]{Matthews}
Matthews, J.M., Kusching, R., Guenther, D.B., Walker, G.A.H., Moffat, A.F.J., Rucinski, S.M., Sasselov, D., Weiss, W.W. 2004, Nature, 430, 51

% \bibitem[Matthews et al. (in preparation)]{MatthewsHD}
% Matthews, J.M. et al. 2007, \apj, in preparation

\bibitem[Melo et al.(2006)]{Melo}
Melo, C., Santos, N.C., Pont, F., Guillot, T., Israelian, G., Mayor, M., Queloz, D., Udry, S. 2006, \aap, 460, 251

\bibitem[McLaughlin (1924)]{McLaughlin}
McLaughlin, D.B. 1924, \apj, 60, 22

\bibitem[Miller-Ricci et al.(2007a)]{Miller}
Miller-Ricci, E., Rowe, J.F., Sasselov, D., Matthews, J.M., Guenther, D.B., Kuschnig, R., Moffat, A.F.J., Rucinski, S.M., Walker, G.A.H., Weiss, W.W.
2007, \apj, submitted 17 Oct 2006 

\bibitem[Miller-Ricci et al. (2007b)]{MillerTwo}
Miller-Ricci, E., Rowe, J.F., Sasselov, D., Matthews, J.M., Guenther, D.B., Kuschnig, R., Moffat, A.F.J., Rucinski, S.M., Walker, G.A.H., Weiss, W.W.
2007, \apj, submitted 12 Apr 2007 

\bibitem[Papaloizou \& Szuszkiewicz (2005)]{Papaloizou}
Papaloizou, J.C.B., Szuszkiewicz, E. 2005, \mnras, 363, 153

\bibitem[Pont et al.(2006)]{Pont}
Pont, F. et al. 2006, \mnras, 373, 231

\bibitem[Pont et al.(submitted)]{PontPrep}
Pont, F. et al. 2007, \aap, astro-ph/0707.1940

\bibitem[Press et al.(1992)]{Press}
Press, W.H., Flannery, B., Teukolsky, S.A., Vetterling, W.T. 1992. Numerical Recipes in Fortran, 2nd ed. (Cambridge University Press, Cambridge, England).

\bibitem[Raymond et al.(2006)]{Raymond}
Raymond, S., Mandell, A., \& Sigurdsson, S. 2006, Science, 313, 1413

\bibitem[Rossiter (1924)]{Rossiter}
Rossiter, R.A. 1924, \apj, 60, 15

\bibitem[Rowe et al.(2006a)]{Rowe}
Rowe, J.F., Matthews, J.M., Seager, S., Kuschnig, R., Guenther, D.B., Moffat, A.F.J., Rucinski, S.M., Sasselov, D., Walker, G.A.H., Weiss, W.W. 2006, \apj, 646, 1241.

% \bibitem[Rowe et al. (in preparation)]{Rowe2}
% Rowe, J.F., Matthews, J.M., Seager, S., Kuschnig, R., Guenther, D.B., Moffat, A.F.J., Rucinski, S.M., Sasselov, D., Walker, G.A.H., Weiss, W.W. 2006b, \apj, in preparation.

\bibitem[Sahu et al.(2006)]{Sahu}
Sahu, K.C., Casertano, S., Bond, H.E., Valenti, J., Smith, T.E., Minniti, D., Zoccali, M., Livio, M., Panagia, N., Piskunov, N., Brown, T.M., Brown, T., Renzini, A., Rich, R.M., Clarkson, W., Lubow, S. 2006, Nature, 443, 5

\bibitem[Steffen \& Agol (2005)]{Steffen}
Steffen, J.H., Agol, E. 2005, \mnras, 364, L96

\bibitem[Thommes (2005)]{Thommes}
Thommes, E. 2005, \apj, 626, 1033

\bibitem[Valencia et al.(2006)]{ValenciaA}
Valencia, D., O'Connell, R., \& Sasselov, D. 2006a, Icarus, 181, 545

\bibitem[Valencia et al.(2007)]{ValenciaB}
Valencia, D., Sasselov, D., \& O'Connell, R. 2007, \apj, 656, 545

\bibitem[Walker et al.(2003)]{Walker}
Walker, G. A. H., et al. 2003, \pasp, 115, 1023

\bibitem[Winn et al.(2006)]{WinnLambda}
Winn, J.~N., Johnson, J.~A., Marcy, G.~W., Butler, R.~P., Vogt, S.~S., Henry, G.~W., Roussanova, A.,
Holman, M.~J., Enya, K., Narita, N., Suto, Y., Turner, E.~L. 2006, astro-ph/0609506

\bibitem[Winn et al.(2007)]{WinnPsi}
Winn, J.~N., Holman, M.J., Henry, G.W., Roussanova, A., Enya, K., Yoshii, Y., Shporer, A., Mazeh, T., Johnson, J.A., Narita, N., Suto, Y. 2007, \aj, astro-ph/0612224

\bibitem[Wright et al.(2004)]{Wright}
Wright, J.~T., Marcy, G.~W., Butler, R.~P., Vogt, S.~S. 2004, \apjs, 152, 261

\bibitem[Zhou et al.(2005)]{Zhou}
Zhou, J.L., Aarseth, S.J., Lin, D.N.C., Nagasawa, M. 2005, \apjl, 631, L85

\end{thebibliography}
\end{document}